\documentclass[preprint2]{aastex}

\usepackage{soul}
\usepackage{color}

\slugcomment{}

\shorttitle{Accretion and thermohaline convection in exoplanets-host stars}
\shortauthors{Th\'eado \& Vauclair}

\begin{document}

\title{Metal-rich accretion and thermohaline instabilities in exoplanets-host stars: consequences on the light elements abundances}

\author{Sylvie Th\'eado and Sylvie Vauclair\altaffilmark{1}}
\affil{Institut de Recherches en Astrophysique et Plan\'etologie, Universit\'e de Toulouse, CNRS, 14 avenue Edouard Belin, 31400 Toulouse, France}
\email{stheado@ast.obs-mip.fr}

%\affil{Institut Universitaire de France}
\altaffiltext{1}{Member of Institut Universitaire de France}

\begin{abstract}
The early evolution of planetary systems is expected to depend on various periods of disk matter accretion onto the central star, which may include the accretion of metal-rich matter after the star settles on the main sequence. When this happens, the accreted material is rapidly mixed within the surface convective zone and induces an inverse mean-molecular-weight gradient, unstable for thermohaline convection. The induced mixing, which dilutes the metal excess, may also have important consequences on the light elements abundances. We model and analyze this process, and present the results according to various possible accretion scenarios. We give a detailed discussion of the different ways of treating thermohaline mixing, as proposed by previous authors, and we converge on a consistent view, including the most recent numerical simulations. We show how the observations of light elements in stars can be used as tracers of such events.

\end{abstract}

%% Keywords should appear after the \end{abstract} command. The uncommented
%% example has been keyed in ApJ style. See the instructions to authors
%% for the journal to which you are submitting your paper to determine
%% what keyword punctuation is appropriate.

\keywords{stars:abundances, stars: solar-type, stars: convection, exoplanets, planetary systems}

%% From the front matter, we move on to the body of the paper.
%% In the first two sections, notice the use of the natbib \citep
%% and \citet commands to identify citations.  The citations are
%% tied to the reference list via symbolic KEYs. The KEY corresponds
%% to the KEY in the \bibitem in the reference list below. We have
%% chosen the first three characters of the first author's name plus
%% the last two numeral of the year of publication as our KEY for
%% each reference.

\section{Introduction}

Solar-type stars are currently divided into “exoplanets-host stars” (EHS), which are known to harbor planets, and “non-exoplanets-host stars” (NEHS), for which no planet has yet been detected with the available techniques and instruments. This dichotomy is obviously related to exoplanet detection bias, as stars without detected planets may very well host planets in the presently non-detectable range. In this framework, the studies concerning the observed differences between these two categories of solar-type stars must be taken with caution. However, taking this restriction into account, these studies give interesting information, at least on the observed EHS.

Here we present a theoretical study concerning the fate of the metal-rich (planetary) material which may fall onto EHS during their early lives, and we analyze the consequences of such events on the abundances of light elements. We compare our results with the presently available observations, and give some predictions and discussion for future work.

The observed EHS are, on average, metal-rich compared to NEHS \citep{Santos01}, although some very metal poor EHS have also been observed \citep{Setiawan10}. This average metal excess was a subject of discussion during the last decade, as the star could either has started with a higher metal content than average, or be related to metal-rich accretion onto the star. This last explanation is now ruled out for several reasons including the fact that the metal-rich accreted matter cannot remain inside the outer convective zone but falls down inside the star due to thermohaline convection. \cite{Vauclair04} gave a first discussion of this specific process. She showed how the accreted matter was mixed down inside the star and left only a very small $\mu$-gradient at the end of the mixing process. It is now proved that this remaining $\mu$-gradient is too small to account for the observed overmetallicity. Contrary to \cite{Vauclair04}, \cite{Theado10} suggested that the induced mixing region could go down to the lithium destruction zone. Independently, \citet{Garaud11} performed precise computations of thermohaline convection in a static $1.4 M_{\odot}$ model, with the assumption of accretion of one Jupiter planet, using the thermohaline coefficient derived by \citet{Traxler11} (see below, section 5.7). She confirmed that the remaining metal excess was much too small to account for the observed overabundances, and found that the mixing region extended below the lithium burning region.

The efficiency of the thermohaline mixing and the resulting abundances strongly depend on the atomic diffusion processes at work in the stars. For example in the case of Carbon Enhanced Metal Poor Stars (CEMP), which accrete matter while they have already spent some time on the main sequence, the thermohaline mixing is strongly limited by the pre-existing stable $\mu$-gradient, induced by helium settling \citep{Thompson08}. This is not expected to occur in the case of EHS which accrete matter at the very beginning of their lives, when atomic diffusion did not yet have time to built important helium gradients. However, as will be seen below, even in this case atomic diffusion is not negligible and has to be taken into account.

We precisely analyze the consequences of metal-rich accretion onto solar-type stars, for various stellar masses and various assumptions of accreted masses. We simulate the induced thermohaline mixing using various prescriptions for the mixing coefficient, including the most recent 2D simulations by \cite{Denissenkov10} and 3D simulations by \citet{Traxler11} and we compare the results. We show that the situation may be quite complex. In any case, if some metal-rich layer can remain in the stellar outer regions, it is much too small to be observable. This confirms the primordial origin of the average overmetallicity observed in EHS. On the other hand, lithium and beryllium destruction layers are reached in most cases and these elements may be partially destroyed.

In the present computations, we do not add any other mixing process which can also destroy lithium and beryllium, like rotation-induced mixing or mixing by internal waves. The reason is that we want to disentangle the various processes acting on the light elements and look at the effects of thermohaline convection only. In the future, these other mixing processes should be added for a complete description of light element evolution in solar-type stars.

In section 2, we give a review of the possible accretion phenomena expected on EHS and their observed evidences; we also discuss in this section the induced thermohaline process. A review of the observed element abundances in EHS versus NEHS stars is given in section 3. Section 4 is devoted to the modeling techniques. Section 5 presents our computations and results; the effects of the various parameters like stellar mass, accretion mass, different prescriptions for the mixing, etc. are compared in detail in this section. Finally the consequences of these results are discussed in section 6, followed by the conclusion.

\section{The accretion phenomenon and the induced-instabilities}
\subsection{Accretion onto exoplanet host stars}
\subsubsection{The planetary system evolution}
It is now well-established that stars may experience episodic accretion along their evolution through different processes.
The planet formation period which extends from the star birth to the end of the PMS phase \citep{Ida08} is a favorable epoch for such events. 
The interactions of a young planet with its surrounding disk affect the planet's orbital energy and angular momentum. As a consequence the planet may migrate radially towards or away from the central star \citep{Terquem10,Lubow10}.

At the end of the protoplanetary disk phase, the planetary system architecture is not necessarily set. Current models for planetary system evolution indicate that various instabilities may commonly arise within the first gigayear after the system birth, as it moves towards a stable configuration. During this phase, the system experiences quasi-stable periods interrupted by more or less violent restructuring episodes. These events which may result in orbital reconfigurations, collisions, gravitational captures or ejections, may be favoured and intensified by the presence of asteroids and debris belts. During these catastrophic events, planetesimals, asteroids or debris may be pushed towards the central region and fall onto the inner objects and/or the central star.

\subsubsection{The Solar case evidences}
Evidences for the previously described evolution are present in our Solar System and comes from a variety of sources. 

Planetary formation theories suggest that the giant planets formed on circular and coplanar orbits different from their current path \citep{Tsiganis05}. 
The asteroid and Kuiper belts are believed to have been formed with considerably more mass than their present content \citep{Stern96,Obrien07}, their radial location may have changed over time \citep{Gomes05,Thommes07} and the dynamical structure of the Kuiper belt provides evidence for sculpting by planet migration \citep{Chiang07,Morbidelli08}.
The petrology and cratering records on the Moon and on telluric planets suggest that a cataclysmic spike in the cratering rate occurred nearly 700 million years after the planets formed \citep{Hartmann00}. This event which is referred to as the Late Heavy Bombardment (LHB), is characterized by an intense period of planetesimal bombardment probably originated from the existing asteroid belt \citep{Kring02,Tagle08} and spreading throughout the inner Solar System \citep{Strom05}. This episode would have last 50 to 200 million years. 

Models and theories have been proposed to explain these observations \citep{Tsiganis05,Thommes07,Chiang07,Ford07,Thommes99,Gomes05,Morbidelli05,Tsiganis05}. According to the most popular ones, at the end of the protoplanetary disk phase, the gas giant planets were surrounded by a rich and interacting asteroid belt imposing slow planet migrations. As a consequence of these migrations, the occurrence of a 1:2 orbital resonance reached by Jupiter and Saturn would have been the catalyst for a catastrophic episode. This phenomenon would have lead to a dramatic destabilization and a rapid restructuring of the outer Solar System, involving the giant planets and the asteroid belt. As a result of this process, the giant planets would have evolved to their current orbits while most of the components of the asteroid belt would have been ejected from the planetary system or simply pushed towards outer regions (forming the current Kuiper belt). However some objects of the belt would have experiment high orbital eccentricity motions allowing them to enter the inner Solar System and impact with telluric objects, resulting in the LHB \citep{Gomes05}. 

The solar system history probably reveals a typical scenario for planetary system evolution. Various models indicate that the instabilities encountered by the Solar System may be a common occurrence for planetary systems and that this can occur after up to 1 Gyr of quasi stable evolution. 

\subsubsection{Interference by asteroid belts}
The presence of solid bodies belts may interfere in the stabilization process of  planetary systems. In particular, as stated in the previous section for the Solar System case, perturbations of asteroid and/or debris disks may result in the scattering of planetesimals throughout the planetary system, favouring the occurrence of intense impact episodes. As most stars are supposed to form surrounded by protoplanetary disks, they are expected to present remnant belts (seen as asteroids or debris disks) resulting from a partial clearing process of their original disks. Indeed massive circumstellar disks are observed around several hundred main sequence stars and preferentially around exoplanet host stars (hereafter EHS) \citep{Bryden09,Wyatt07}.
We consequently expect planetary systems to offer preferential conditions for violent bombardment episodes and consequently for accretion event onto the central star.

\subsection{The thermohaline instability}
The accretion of planetary (metal rich) material onto a star produces a brutal increase of its surface metal content. The metal-rich accreted material is rapidly mixed within the surface convective zone: a metal-enriched zone may then overlay lighter layers which produces an unstable weight gradient in the external layers of the star. A similar situation is often encountered in oceanography, when warm salted water lies upon cool unsalted one. Large inverse salinity gradients make warm salted blobs diffuse rapidly downward in spite of the stabilizing temperature gradient. Heat diffusing more rapidly than salt, the warm salted blob falling down in cool fresh water experiences a temperature decrease before the salt has time to diffuse out; the blob continues falling because of its weight, while unsalted matter rises around. This special mixing process which takes the form of elongated cells, the ``salt fingers", is called ``thermohaline convection". It proceeds until the fingers mix with the surroundings and a stable salinity gradient is restored. 

A similar type of convection can grow inside stars. It occurs in a stable stratification that satisfies the Ledoux criterion for convective stability: 
\begin{equation}
 \nabla_{ad}-\nabla_{rad}+\frac{\phi}{\delta}\nabla_{\mu}>0
\end{equation}
but where the molecular weight decreases with depth:
\begin{equation}
\nabla_{\mu}=\frac{d\ln \mu}{d \ln P}<0
\end{equation}
With the classical notations, $\nabla_{ad}$ and $\nabla$ are respectively the usual adiabatic and local gradients ($d\ln T/d \ln P$), $\displaystyle \phi=(\partial \ln \rho / \partial \ln \mu)_{P,T}$ and $\displaystyle \delta= (\partial \ln \rho / \partial \ln T)_{P,\mu}$. 

In the stellar case, the role of the aquatic salinity is then played by the molecular weight. When a blob of heavy matter (i.e with high molecular weight) begins to fall down, heat is exchanged with the surroundings more rapidly than particles so that it goes on falling like a finger, forcing light matter to go up instead. This happens every time heavy element accumulation occurs above lighter layers. It leads to short episodes of deep thermohaline mixing which proceed until the molecular-weight gradient becomes too small to trigger the instability. This can be summarized with the following condition (see \cite{Vauclair04} and \cite{Traxler11}):

\begin{equation}
1 < R_0 < \frac{1}{\tau}
\end{equation}

where $R_0$ represents the ratio:
\begin{equation}
R_0 = \frac{\nabla_{ad}-\nabla_{rad}}{\nabla_{\mu}}
\end{equation}
and $\tau$ is the inverse Lewis number, ratio of the molecular diffusivity to the thermal diffusivity.

A more complete description of the methods used to model thermohaline convection is given in Section 4.2.

Thermohaline instabilities due to inverse $\mu$-gradients are expected to arise in stellar interiors in a variety of already identified contexts. This is the case for instance when helium or carbon rich material is deposited at the surface of a star in a mass-transferring binary \citep{Stothers69,Stancliffe07,Thompson08}. It also develops in stars where radiative levitation produces heavy element accumulation regions \citep{Theado09}. The thermohaline mixing has also been proposed as a possible mechanism to account for the photospheric composition of low and intermediate mass giant \citep{Eggleton06,Eggleton08,Charbonnel07,Cantiello08,Siess09,Charbonnel10,Stancliffe10}. In such stars an inverse $\mu$-gradient is created by the $^3$He($^3$He,2p)$^4$He reaction in the external wing of the hydrogen burning shell. An inverse $\mu$-gradient may also arise after the ignition of $^4$He burning in a degenerate shell (i.e. core helium flash; see \cite{Thomas67,Thomas70}).

Up to now the effects of accretion-induced thermohaline instabilities on EHS have been little studied. They have been discussed by \cite{Vauclair04} and more recently by \citet{Garaud11}, but no detailed stellar evolutionary models have yet been computed to investigate the effects of accretion and thermohaline convection on the stellar lithium, which is the aim of the present paper.

\section{Observational background : abundances in exoplanet host stars}
\subsection{The overmetallicity of EHS}
Stars hosting planets compared with other stars of the same spectral types clearly show a metallicity excess of a factor 2 on average. Two scenarios have been proposed to explain the metal enrichment of EHS. The first one, the accretion hypothesis, considers that EHS formed with normal abundances but experienced during the planet formation period metal-rich material accretion \citep{Gonzalez98,Murray01}. The ingestion of planets, planetesimals, asteroids, comets, dust would then produce a local surface overmetallicity. The second scenario, the primordial hypothesis, assumes that central stars of planetary system form out from metal rich molecular clouds \citep{Pinsonneault01,Santos01,Santos03} : the overmetallicity of the protostellar gas would be a necessary condition for planets to form around stars. In this case, the stars would be metal-rich from the center to their surface. 

Several studies have tried to differentiate between the two possible explanations using different techniques including asteroseismology \citep[e.g.][]{Bazot04}.
Today a consensus seems to be achieved which favors the primordial hypothesis. Several arguments have been invoked which rule out the accretion scenario. It was first thought that the dilution of metal-rich material in stellar external convective zones should lead to metallicity variations depending on the stellar mass, which were not observed \citep{Pinsonneault01,Santos01}. Another argument was that the overmetallicity observed in EHS would require the ingestion of several tens of earth, which is not completely excluded but seems quite high \citep{Israelian04}. More recently, as already discussed in previous section, it was recognized that the overmetallic matter diluted in the surface convective zone would rapidly sink in the radiative interior due to thermohaline instability effects, so that only a small fraction of the accreted matter could remain in the outer layers (see \cite{Vauclair04} and \citet{Garaud11}).

\subsection{The lithium and beryllium enigma}
\subsubsection{Observational context}
The potential discrepancy between lithium abundances in EHS and NEHS is a controversial subject which has divided the observers community for more than 10 years. \citet{Gonzalez00} first suggested that EHS present smaller lithium abundances than field stars. They were soon contradict by \citet{Ryan00} and later by \citet{Gonzalez01} who stated that lithium abundances of stars with detected exoplanets are indistinguishable from those of stars without detected planets. \citet{Israelian04}, who revisited this topic, reported a lithium deficiency in EHS within the effective temperature range 5600K-5850K. Their results were confirmed by \citet{Takeda05} and \citet{Chen06} but contradict by \citet{Luck06}. More recently \citet{Takeda07}, \cite{Gonzalez08} and \cite{Israelian09} confirmed an enhanced lithium depletion in Sun like stars with orbiting planets but once more these results were contradict by \cite{Baumann10} who attributed the lithium depletion derived by \cite{Israelian04,Israelian09} to a wrong interpretation of observational data or to observational bias. 

The situation is still more difficult for beryllium. \citet{Delgado11} and \citet{Galvez11} have determined the beryllium abundances in a sample of solar-type stars and tried to determine a difference between exoplanet-host stars and stars without detected planets. They may find an extra beryllium depletion for the coolest stars (T$_{eff}$$\rm <5500$ K), but this result still has to be confirmed.

\subsubsection{Theoretical context}
Considering the potential lithium deficiency observed in EHS as true, theoreticians proposed several explanations to account for this ``extra'' depletion. Some of them suggest that the lithium destruction and the planet formation both results from special conditions surrounding some stars (e.g. overmetallicity, long-lived disk). Others consider that the destruction is directly related to the presence of planets.

\paragraph{Metallicity effects.}
\cite{Castro09} computed stellar evolutionary models including atomic diffusion and rotation-induced mixing. In this framework they investigated the effects of various primordial overmetallicities on the lithium depletion. They showed that the EHS-lithium deficiency cannot be attributed to overmetallicity effects : extra-mixing is required below the surface convective zone to account for the Li observations.

\paragraph{Stellar rotational history.}
\citet{Bouvier08} investigated a possible connection between a lithium deficiency, the presence of planets and the rotational history of solar type stars. Thanks to rotational evolution models, he showed that contrary to ZAMS fast rotating stars, the slow ZAMS rotators develop strong differential rotation between their surface convective zone and their radiative interior, which leads to hydrodynamical instabilities responsible for enhanced lithium depletion. He concluded that EHS were slow rotators on the ZAMS and that the protoplanetary long-lived disk probably responsible for their slow rotation rate could also be a necessary condition for the planet formation. This could explain the correlation observed between the presence of planets and large lithium destructions.   

\paragraph{Formation, migration and accretion of planets.}
Several authors suggest that the strong lithium depletion observed in EHS could be directly related to the mechanism of formation or migration of planets.  
This idea is supported by observations from \cite{Chen06} which show that overmetallic stars without detected planet do not present strong lithium depletion.

\cite{Israelian04} suggest that protoplanetary disks lock a large amount of angular momentum during the planet formation mechanism thus inducing some rotational braking in the host star during the PMS. The mixing resulting from the induced differential rotation could lead to a large lithium depletion.
\cite{Israelian04} also suggests that the lithium destruction could be associated with a late migration of giant planets at the end of the PMS. Angular momentum transfer from the planetary disk to the outer layers of the star and/or tidal effect of the giant planet on the host star could induce differential rotation and then mixing in the host stars. 

If the migration process is overly efficient, planets may be accreted onto the star. \citet{Montalban02} investigate the effects of the accretion of Jupiter-like planets on the lithium isotopic abundances in main sequence stars, using either standard stellar models or models with a light mixing process induced by gravity waves. They do not take into account any thermohaline convection, so that the stellar outer layers are not mixed with the lithium depleted regions. As a consequence, they obviously find an increase of the lithium abundance and the 6/7 isotopic ratio due to the planet chemical composition. Baraffe \& Chabrier (2010) explores the effects of early episodic accretion (i.e. occurring during the PMS) on the internal structure of young (fully or nearly fully convective) low mass and solar type stars. They conclude that an early protostar accretion history can affect the internal structure of the star and can severely enhanced the Li-depletion in these object.

\section{Modeling}
We computed stellar models of solar type stars, using the Toulouse-Geneva Evolution Code (TGEC). The code is described in detail in \citet{Richard04} and \citet{Hui08}. All our models include atomic diffusion computed as described in \citet{Hui08} and \citet{Theado09}. The selective diffusion of a large number of individual elements important for the stellar structure is computed with respect to hydrogen, including the effects of the local pressure gradient, temperature gradient, concentration gradient and radiative transfer. At each time step, all the abundances are renormalized to account for the correlated motion of hydrogen and the induced modifications of the opacities are taken into account. When macroscopic motions occur, which can be described in terms of a macroscopic diffusion coefficient, the mixing effect is added as a new term in the diffusion equation, acting on the abundance gradient, as described in many papers (e.g. \citet{Vauclair78}, \citet{Vauclair82}, \citet{Richard96}, \citet{Vauclair03}, \citet{Michaud05}, \citet{Korn06}, etc.). 

In the TGEC code the main isotopes of the following elements are treated separately: He, Li, Be, B, C, N, O, Ne, Mg, Ca, Fe. The other species are treated together. For each separate isotope i, the diffusion equation is written:
\begin{equation}
\frac{\partial (\rho c_i)}{\partial t}+ div (\rho c_i v_i)=0
\end{equation}
where c$_i$ is the concentration of the isotope and $(\rho c_i v_i)$ the diffusion flux. In atomic diffusion theories, $v_i$ is traditionally called the ``diffusion velocity''. The expression is deduced from detailed theories of transport processes in multicomponent non-uniform gases \citep[e.g.][]{Chapman70}. For test atoms, it may be written:
\begin{eqnarray}
v_i=D_i \biggl(-\frac{D_i+D_{th}}{D_i} \nabla \ln c + k_P \nabla \ln P \nonumber \\ + k_T \nabla \ln T + \frac{m_i g_{i}}{k T} \biggr)
\end{eqnarray}
where the first term is the classical diffusion term and the three other ones are respectively the pressure, thermal and radiative diffusion terms \citep[e.g. Eq. 2 of][]{Theado03a}. D$_i$ is the atomic diffusion coefficient and D$_{th}$ the mixing diffusion coefficient. In the cases where atomic diffusion is negligible, these equations reduce to the classical diffusion (mixing) equation.
%\begin{equation}
%\frac{\partial (\rho c)}{\partial t}=div (\rho D_{th} grad c)
%\end{equation}

In the following, we discuss how accretion and thermohaline convection are introduced in the code.

\subsection{Accretion}
\label{sectionaccr}
We assume that stars could swallow planetary material during their early main sequence life. The accretion of a given amount of matter is modeled through an increase of the metal content of the surface convective region. The accreted mass is expressed in Jupiter Mass (hereafter $\rm M_{Jup}$) or Earth Mass unit (hereafter $\rm M_{\oplus}$); its composition is assumed similar to the solar mixture determined by \citet{Grevesse93} but devoid of hydrogen and helium. The composition of the surface convective zone is computed again after each accretion episode : the mass fractions of the chemical elements are computed taking into account the newly accreted matter but the structure of the star is supposed unaltered by the accreted material. 

In the case of large amounts of accreted matter, the $\mu$-gradient is large enough at the beginning to destabilize the medium by dynamical convection. This is modelled by a very large diffusion coefficient simulating complete mixing. Thermohaline mixing begins when the $\mu$-gradient is small enough for the medium to be stable against dynamical convection. The prescriptions used to describe this mixing are discussed below.

\subsection{Thermohaline convection}
\label{modelthermo}
The effects of thermohaline convection as a mixing process in stars are far from trivial.
\citet{Ulrich72} and \citet{Kippenhahn80} (hereafter referred to as KRT) proposed to model the thermohaline mixing process in terms of a diffusion coefficient proportional to the inverse $\mu$-gradient:
\begin{equation}
D_{th}=C_t \frac{4 a c T^3}{3 c_p \kappa \rho^2} \frac{H_p}{\nabla_{ad}-\nabla} |\frac{d \ln \mu}{dr}|
\label{dth}
\end{equation}
which can simply be written:
\begin{equation}
D_{th}=C_t \kappa_T R_0^{-1}
\end{equation}

where $\kappa_T$ is thermal diffusivity, $ H_{P}$ is the pressure scale height, $a$ the radiation constant, $c$ the speed of light, $T$ the temperature, $\nabla_{ad}$ the adiabatic gradient, $ c_P$ the specific heat at constant pressure, $\kappa$ the opacity. The variable $C_t$ is a factor involving the aspect ratio (length/width) of the fingers. KRT and \cite{Ulrich72} considered different geometries for the fingers and, as a consequence, they deduced different $C_t$ values (12 for KRT and 658 for \cite{Ulrich72}). In their computations of red giant stars, \cite{Charbonnel07} considered this parameter as free and adjusted it ($C_t = 1000$) to reproduce the abundance observations in evolved stars. On the other hand \citet{Theado09} used the original KRT value with $C_t = 12$ for the computations of diffusion induced thermohaline convection.

A large step forward has recently been acomplished owing to the numerical simulations performed by \cite{Denissenkov10} (2D) and \citet{Traxler11} (3D). They both deduced from their simulations of fingering convection a near-unit aspect ratio of the fingers much smaller than that used by \citet{Charbonnel07}. In the intermediate regime, their different expressions for the thermohaline mixing coefficients lead to values very close to the KRT ones, if computed with $C_t$ = 12. Important differences appear only for the boundary values of the $R_0$ ratio, when $R_0 \rightarrow 1/\tau$, which corresponds to the case of very small $\mu$-gradients, or $R_0 \rightarrow 1$, which corresponds to the largest possible $\mu$-gradients before dynamical convection. 

The \cite{Denissenkov10} coefficient may be written as:
\begin{equation}
D_{th}=C_t \kappa_T (R_0-1)^{-1} (1-R_0 \tau)
\end{equation}
which reduces to the KRT one for $1<< R_0 <<1/\tau$, but vanishes for $R_0 = 1/\tau$ and goes to infinity when $R_0 = 1$.

\citet{Traxler11} (hereafter TGS) have performed 3D simulations of thermohaline convection with various sets of physical parameters, and found that the ratio of the turbulent to the diffusive fluxes settled onto a single universal profile for all parameter values, which seems a robust result. They derive a completely empirical expression for the mixing coefficient which, at first sight, looks quite different than the previous ones, but present also some similarities:
\begin{equation}
D_{th}=101 \sqrt{ \kappa_{\mu} \nu} \exp{ (-3.6 r)}(1-r)^{1.1}
\end{equation}
where, in stellar conditions for which $\tau \ll 1$, $r = (R_0 - 1) \tau$. After some manipulations, we find that in the intermediate regime this mixing coefficient may be approximated as $D_{th} = C_T \kappa_T \tau$, which is slightly smaller than the KRT one with $C_t$ = 12. Once again, important differences only appear for the behavior in the extreme cases. Here $D_{th}$ vanishes for $R_0 = 1/\tau$ but it reaches a finite value when $R_0 = 1$. This last point will be discussed in section 5.7.
An illustration of this discussion may be found in \ref{coefficients} where we present a graph of the KRT and TGS mixing coefficients precisely computed in one of our models.

In the following, we first present results obtained with the KRT prescription using $C_T$ = 12, and we also do computations with the assumption $C_T$ = 1000 for comparison purposes. We show that in this case, the resulting lithium and beryllium depletions are too high to be realistic, which confirms that this coefficient is overestimated. Then we present computations obtained with the new TGS empirical expression and we compare it with the previously obtained results. 

\section{Computations}
We made computations for stars with masses between 0.70 and 1.30M$_{\odot}$. We tested various initial metal contents with metallicity values ranging from [Fe/H]=-0.20 to +0.20. We computed models with a solar initial helium mass fraction and models with an initial helium abundance varying with metallicity (Y$_0$=Y$_G$) according to the galaxy chemical evolution law \citep{Izotov04}. One to fifteen accretion events were introduced in our models with various accretion rates for a total accreted mass ranging from 1M$_{\oplus}$ to 1.5M$_{\rm Jup}$. We present here a choice of our models, representative for all the computations done.

\subsection{The accretion/thermohaline mixing sequence}
In this section we describe in detail the accretion/thermohaline mixing process, and present results obtained with the KRT coefficient ($C_t$ = 12).

\subsubsection{The case of one accretion episode}
As a first example, we present the result obtained for a 1.10 M$_{\odot}$-model with Y$_0$=Y$_G$=0.29 and [Fe/H]=0.20. This model experiences a single accretion event of 0.03M$_{\rm Jup}$ two million years after its arrival on the ZAMS. This accretion rate is of the order of a giant planet core, which represents a realistic value for the mass of heavy material collected by the star through an impact with a Jupiter like object.

\begin{figure*}
\center
\includegraphics[width=0.5\textwidth,bb=300 450 593 718]{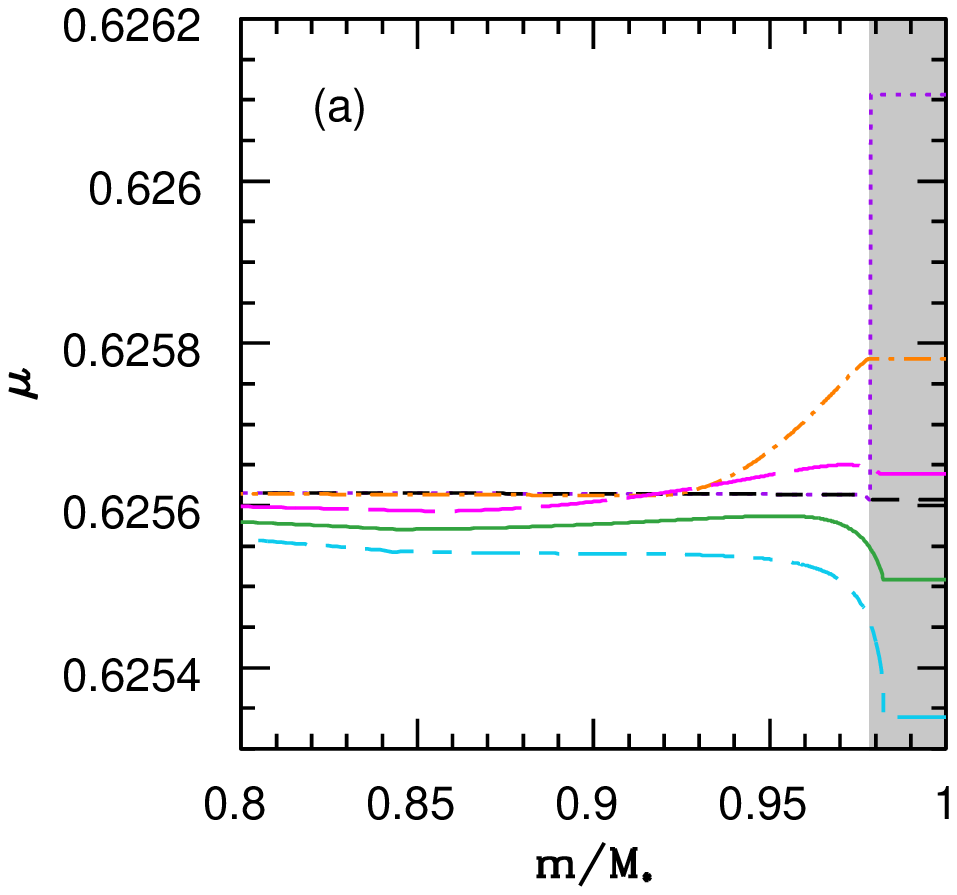}%
\includegraphics[width=0.5\textwidth,bb=300 450 593 718]{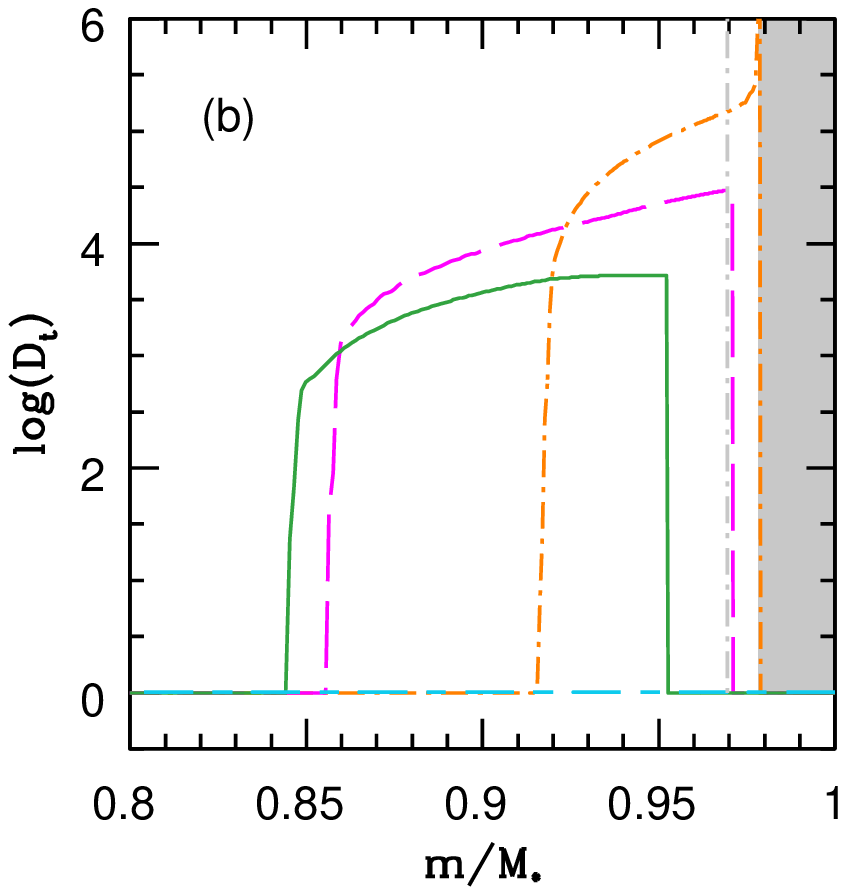}%
\caption{Description of one accretion/mixing period in a 1.10 M$_{\odot}$ model undergoing 0.03M$_{Jup}$ accretion event 2 Myr after its arrival on the ZAMS. Figure a : molecular weight profiles in the external regions of the model. The dashed (black) line displays the atomic diffusion-induced $\mu$-profile at 2 Myr just before the accretion episode, the dotted (purple) line represents the theoretical $\mu$-profile resulting from the instantaneous dilution of the planetesimal in the surface convective zone. The dashed-dotted (orange), long-dashed (magenta), solid (green) and long-dashed dashed (blue) lines represent the $\mu$-profiles 2, 25, 54 and 89 Myr after the accretion event. After 4 Myr, the $\mu$-gradient is still unstable below the convetive zone, whereas after 25 Myr, a diffusion-induced helium gradient leads to a stable $\mu$-gradient in these layers, as discussed in the text. Figure b : profiles of the diffusion coefficient (cm$^2$.s$^{-1}$) modeling the thermohaline mixing (same line convention as in Figure a). The vertical line locate approximately the top of the Li-nuclear depletion zone. The grey regions locate the surface convective zone. After the accretion episode, mixing occurs deeper and deeper and links the outer layers down to the lithium burning region (dashed-dotted, orange line), until the competition with atomic diffusion begins to lead to stable layers below the convective zone (solid, green line). After 89 Myr (long-dashed dashed, blue line), the thermohaline diffusion coefficient is flat and there is no more mixing.}
\label{mudkip1}
\end{figure*}

Figure \ref{mudkip1}a displays the molecular weight profile in the outer regions of the model at various epochs during the ``accretion/mixing period'' (which refers to the period starting with the accretion episode and ending when the thermohaline mixing stops).

The instantaneous dilution of a 0.03M$_{\rm Jup}$ planetesimal in the external convective zone produces an enhancement of the surface metal content of the model. The dotted (purple) line on Figure \ref{mudkip1}a represents the theoretical metallicity increase produced by the dilution in the convective zone of the impacting object. This metallicity increase creates a sharp positive $\mu$-gradient at the transition between the radiative interior and the convective envelope which triggers thermohaline instabilities. The induced mixing acts to soften the unstable $\mu$-stratification. It occurs simultaneously with the atomic diffusion and in particular the downward migration of helium which also contributes to restore a stable $\mu$-gradient. When a flat $\mu$-gradient is reached below the convective zone, thermohaline convection becomes inefficient and a stable $\mu$-gradient induced by helium settling begins to take over. Thermohaline convection still occurs below this stable region. As will be seen in forthcoming sections, this effect is less important for larger accretion rates (section 5.3). 

Figure \ref{mudkip1}b displays the diffusion coefficient used to parametrized the thermohaline mixing at the same epochs as presented on Figure \ref{mudkip1}a. 
The mixing starts right after the accretion event, when an inverse $\mu$-gradient is created at the transition between the radiative and convective zones. It mixes a localized region restricted to the unstable $\mu$-stratified layers lying below the convective zone. This process smooths down the unstable $\mu$-gradient in the mixed zone, it decreases the diffusion coefficient but it also continuously widens the unstable region. The bottom of the mixed zone deepens until a stable stratification can be reached : the mixing then stops and the diffusion coefficient drops to zero. The phases during which the accretion mixing proceeds have strong consequences for light elements like Li and Be as discussed in the following sections.

\subsection{The multiple accretion event case}
We now examine the case of multiple successive accretion events. We present the results obtained for two 1.10M$_{\odot}$ (with the same $ Y_0$ and [Fe/H] values as previously). In both cases the models experience ten accretion events. The first accretion episode occurs 2 Myr after the arrival on the ZAMS, and the following ones occur successively every 2 Myr. In the first model, the amount of accreted matter is 1M$_{\oplus}$ ($\simeq$0.003M$_{\rm Jup}$) at each episode. In the second model, the amount swallowed each time is 0.03M$_{\rm Jup}$, of the order of a giant planet core, which represents a realistic value for the mass of heavy elements collected by the star through an impact with a Jupiter like object. 

\begin{figure*}
  \centering
  \begin{minipage}[c]{0.5\textwidth}
    \centering
    \includegraphics[width=\textwidth,bb=240 162 565 690]{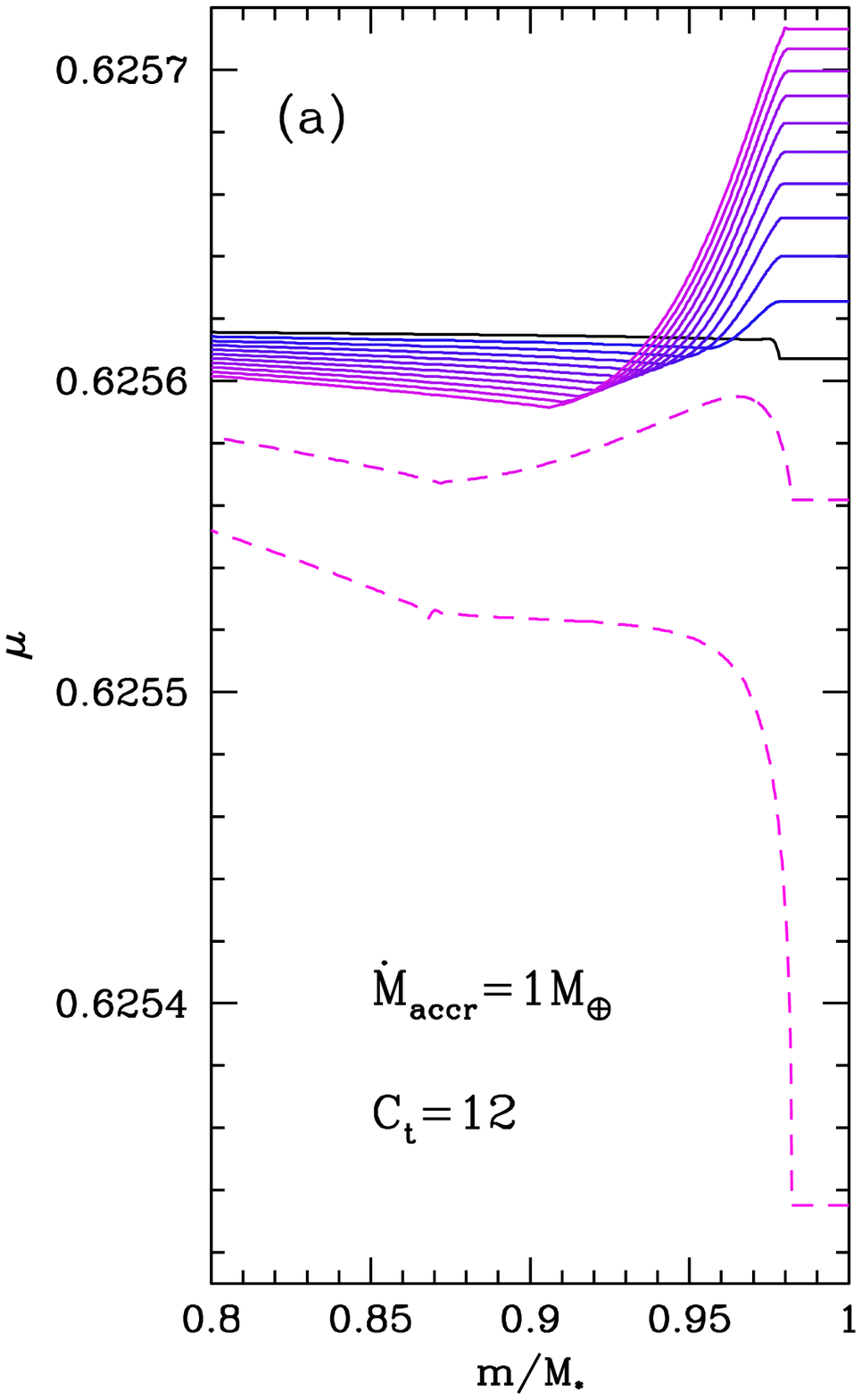} %
  \end{minipage}%
  \begin{minipage}[c]{0.5\textwidth}
    \centering
    \includegraphics[width=0.81\textwidth,bb=315 430 565 690]{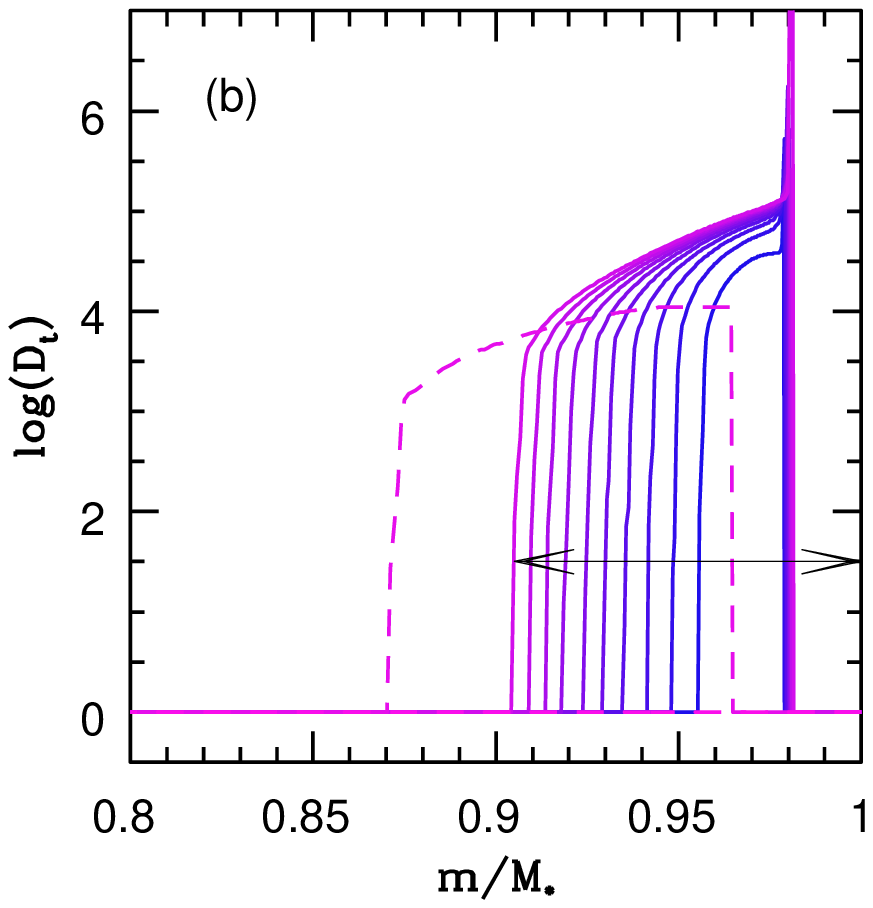} %
    \includegraphics[width=0.81\textwidth,bb=315 430 565 690]{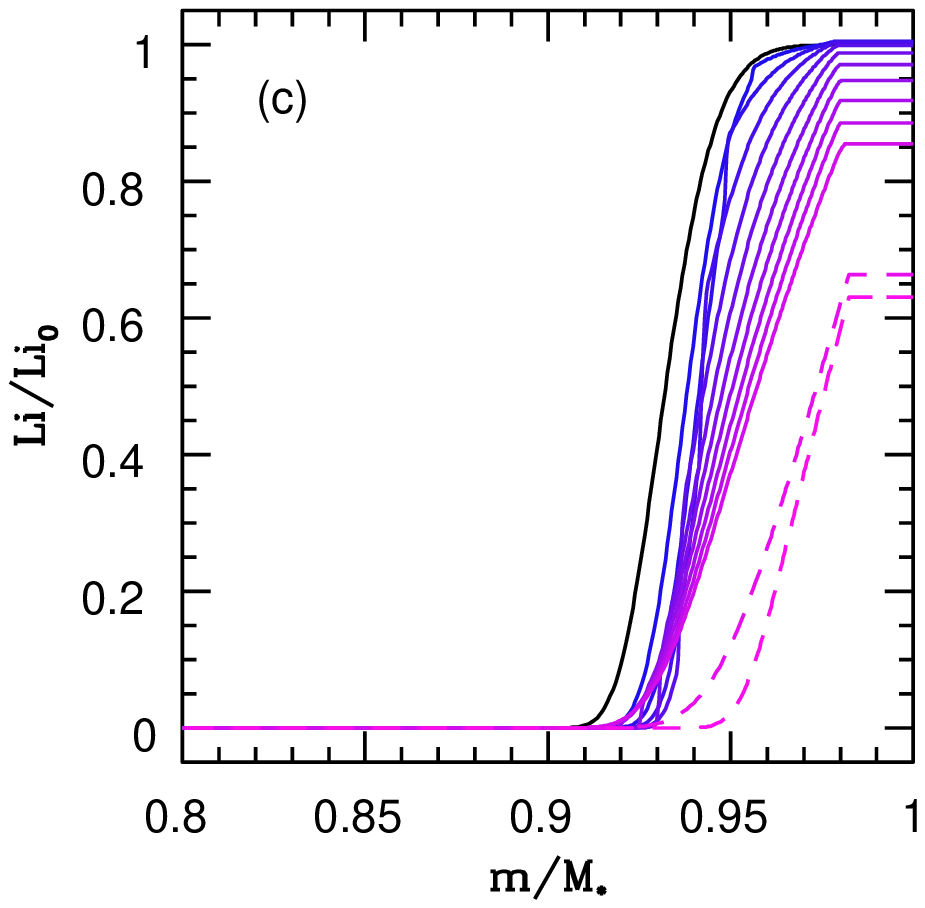} %
  \end{minipage}%
\caption{a) molecular weight, b) thermohaline diffusion coefficient (cm$^2$.s$^{-1}$) and c) Li abundance profiles at various steps during the accretion/mixing process in a 1.10M$_{\odot}$ model experiencing 10 accretion events of 1M$_{\oplus}$. (Li/Li$_0$) represents the ratio between the current lithium abundance and its initial value (i.e. at its arrival on the ZAMS, before accretion and diffusion start). The black lines represent the $\mu$ and Li-abundance profiles at 2Myr on the ZAMS,just before the first accretion event. The dark blue to magenta solid lines present the profiles at 4, 6, 8, 10 ... and 22 Myr.  The light magenta dashed lines show profiles at 51 and 97 Myr (the accretion has stopped but the thermohaline mixing still proceeds). For the two first accretion events, the same process as described in Figure 1 occurs. As an example, the horizontal arrow in the upper right panel locates the mixed region after 10 accretion events.The down right graph show the importance of this process on the surface lithium abundances.}

\label{dkip100a}
\end{figure*}

\begin{figure*}
  \centering
  \begin{minipage}[c]{0.5\textwidth}
    \centering
    \includegraphics[width=\textwidth,bb=240 162 565 690]{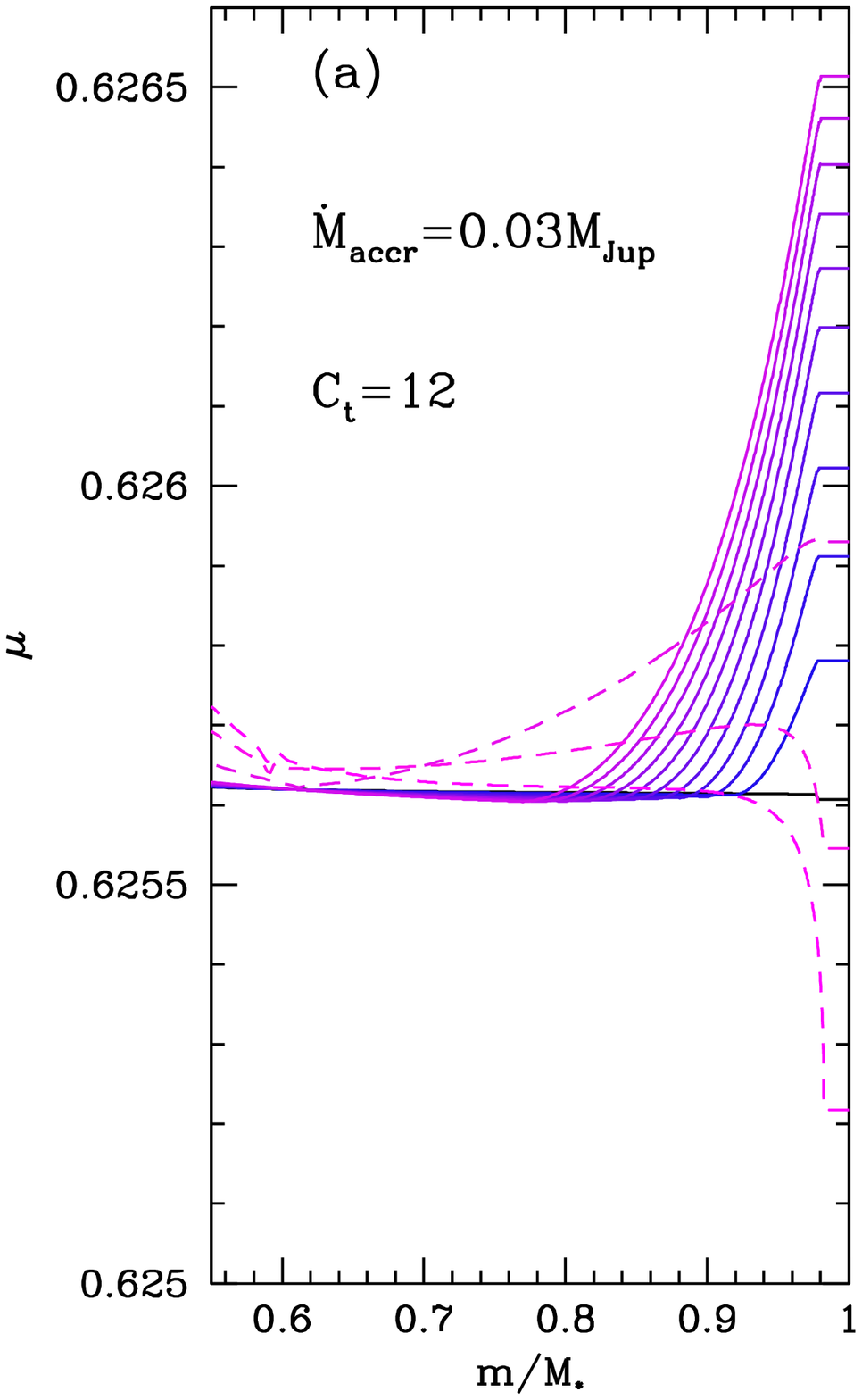} %
  \end{minipage}%
  \begin{minipage}[c]{0.5\textwidth}
    \centering
    \includegraphics[width=0.81\textwidth,bb=315 430 565 690]{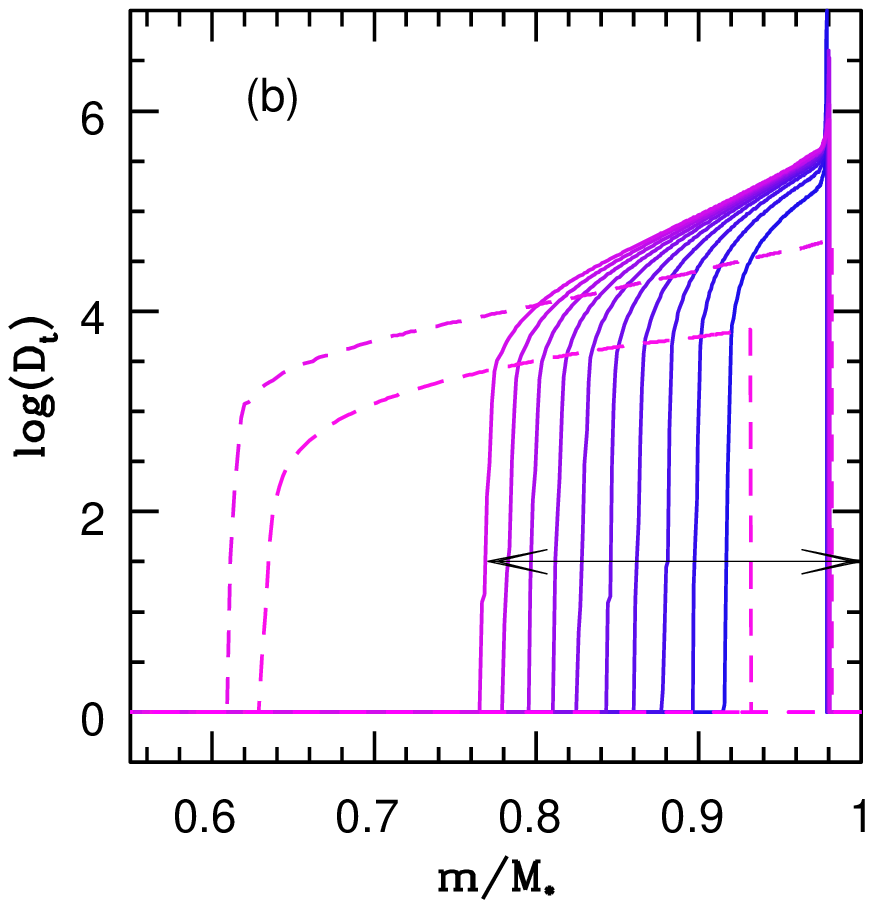} %
    \includegraphics[width=0.81\textwidth,bb=315 430 565 690]{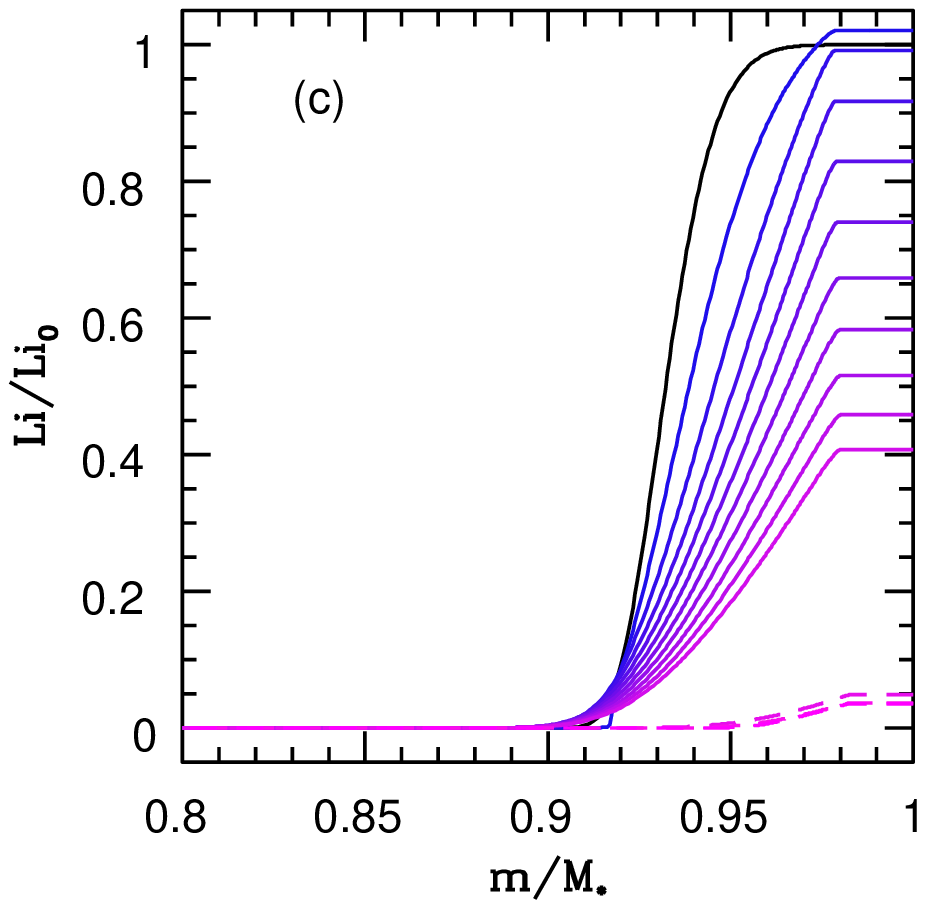} %
  \end{minipage}%
\caption{Same figure as Figure \ref{dkip100a} for a 1.10M$_{\odot}$ model experiencing 10 accretion events of 0.03M$_{ \rm Jup}$. The line convention is the same as in Figure \ref{dkip100a} except for the light magenta dashed lines which show profiles at 74, 167 and 236 Myr on the ZAMS. Here we can see that the small stable helium zone develops after 167 Myr.}
\label{dkip100b}
\end{figure*}

Figure \ref{dkip100a} (a and b) and  Figure \ref{dkip100b} (a and b) present $\mu$ and D$ _t$ profiles inside our models at various steps along the accretion/mixing period. As already discussed, the first accretion event induces a surface $\mu$ increase which triggers the thermohaline instabilities. The resulting mixing tends to smooth down the unstable stratification before the next accretion event occurs. The next successive accretion episodes continuously increase the inverse $\mu$-gradient, which feeds the mixing and increases the width of the mixed zone. After 2 accretion events, the helium settling becomes unable to counteract and freeze the thermohaline mixing before the next event occurs. After the last accretion episode, the thermohaline convection still proceeds, until a stable $\mu$-gradient is restored. During that process, a small stable region develops just below the convective zone. This is the reason why, on the top right graph, one curve, drawn after 51 Myr, present a $D_t$ which vanishes slightly below the convective zone. The time required to restore a stable stratification depends on the accretion rate : the higher the accretion rate, the longer the mixing period and the larger the mixed region. The model experiencing the 1M$_{\oplus}$ accretion rate recovers a stable $\mu$-stratification at 39 Myr (i.e. 37 Myr after the first impact) while the model with the 0.03M$_{\rm Jup}$ accretion rate is stable at 51 Myr (49 Myr after accretion beginning).

As shown on Figures \ref{dkip100a} and \ref{dkip100b}, the top of the Li-depleted region is close to the bottom of the surface convective zone. The thermohaline mixing then allows connecting the stellar surface with the lithium nuclear burning region and is therefore expected to produce a Li depletion. Figures \ref{dkip100a}c and \ref{dkip100b}c represent the lithium profiles inside the two considered models. In both models the first accretion episode does not deplete significantly the lithium surface abundance: the mixing-induced lithium destruction is compensated or even exceeded by the lithium amount brought by the impacting object. The next events increase the unstable $\mu$-gradient, as a result the diffusion coefficient is enhanced and the mixed region extends towards deeper regions : as shown on Figures \ref{dkip100a}c and \ref{dkip100b}c, this leads to large lithium destructions. We will see in the next section that similar features are obtained for Be (the Be depletion is however expected to be smaller than the Li one since its nuclear burning region lies deeper in the stellar interior).
\subsection{The accretion rate effects}

\begin{figure}
\center
\includegraphics[width=0.02\textwidth,bb=17 50 50 550]{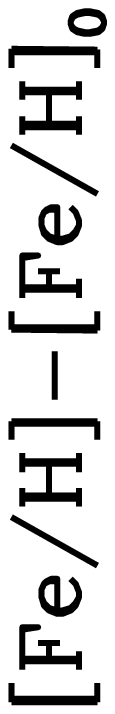}%
\includegraphics[width=0.48\textwidth]{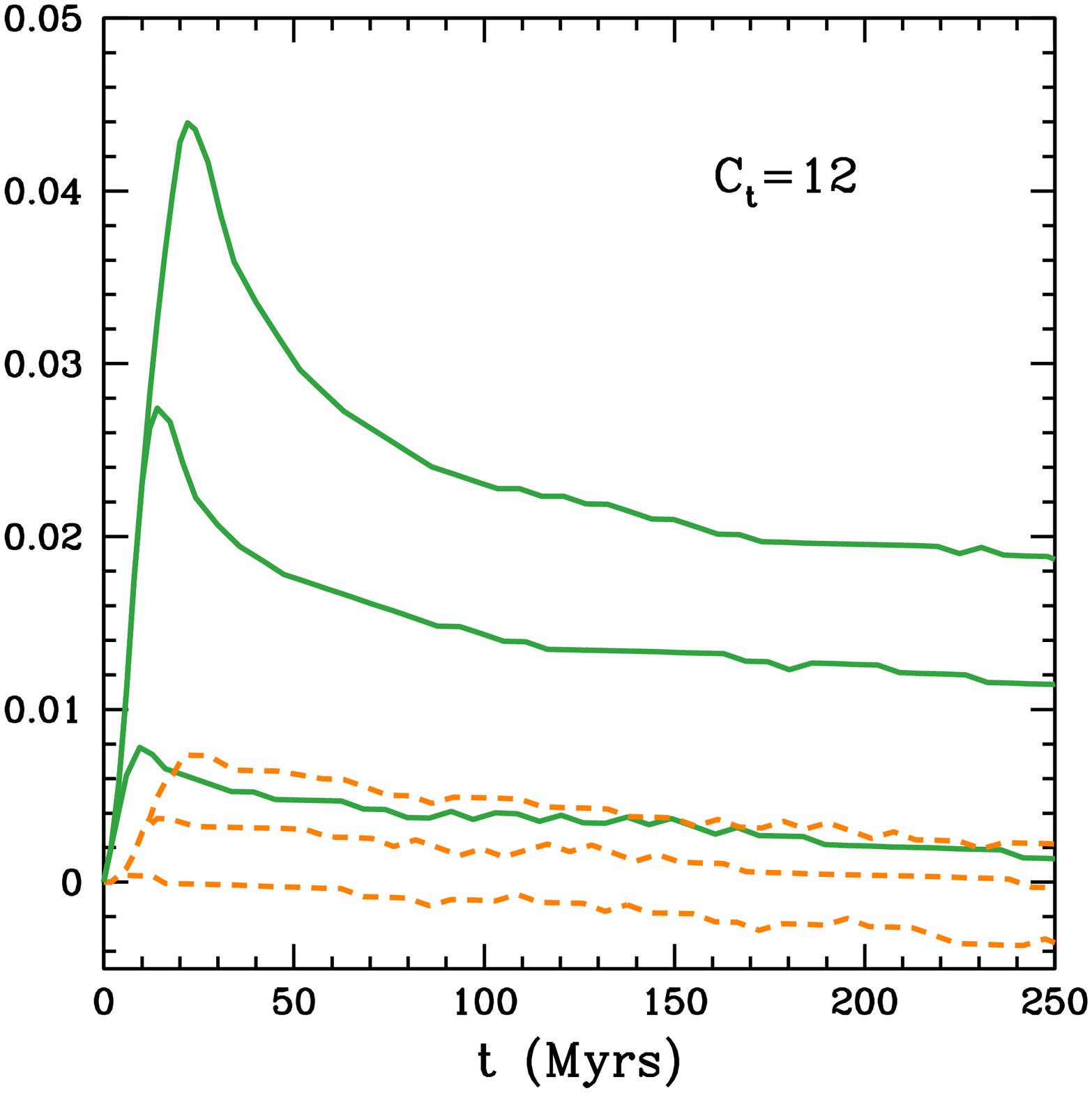}
\caption{Variations of surface overmetallicity along the accretion/mixing period for several 1.10M$_{\odot}$ models. $\Delta [Fe/H]$ is the difference between the current surface metallicity and its initial value. Dashed lines are models experiencing (from bottom to top) : 1, 5 or 10 accretion events of 1M$_{\oplus}$. Solid lines are models experiencing (from bottom to top) : 1, 5 or 10 accretion events of 0.03M$_{\rm Jup}$.}
\label{fesurh}
\end{figure}

In this section, we test the effects of various accretion rates on the surface metallicity and the Li and Be depletions. In this framework, Figure \ref{fesurh} displays the variations of the surface metallicity over the accretion/mixing period for the two series of models (including the two models presented in the previous section). All these models have the same mass (1.10M$_{\odot}$) and initial parameters (in particular the same chemical composition) but they experience various accretion scenarios : 1, 5 or 10 accretion events of 1M$_{\oplus}$ (0.003M$_{\rm Jup}$) for the first series of models and 1, 5 or 10 accretion events of 0.03M$_{\rm Jup}$ for the second series. 

Each accretion episode induces a metallicity increase which obviously depends on the accretion rate. For the considered accretion rates and number of impacts, the overmetallicity does not exceed 0.05 dex. After the end of the accretion period, the overmetallicity decreases under the combined effects of thermohaline mixing and atomic diffusion which makes heavy elements diffuse out of the surface convective zone. When the mixing stops (i.e. before 60 Myr for all the presented models), the remaining metallicity increase is smaller than 0.02 dex and is unable to explain the average overmetallicity of 0.2 dex observed in EHS. 

\begin{figure*}
\center
\includegraphics[width=0.5\textwidth]{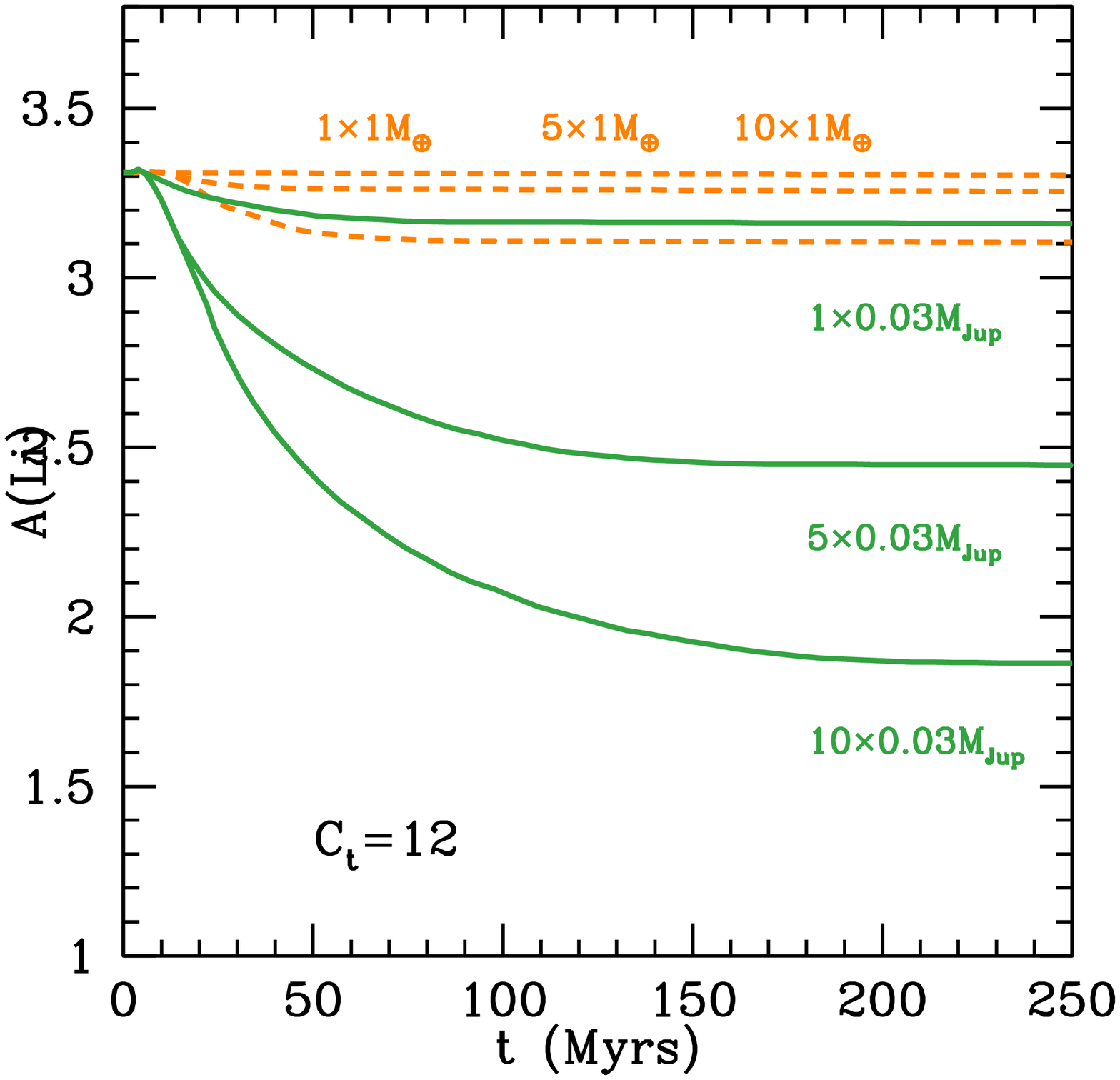}%
\includegraphics[width=0.5\textwidth]{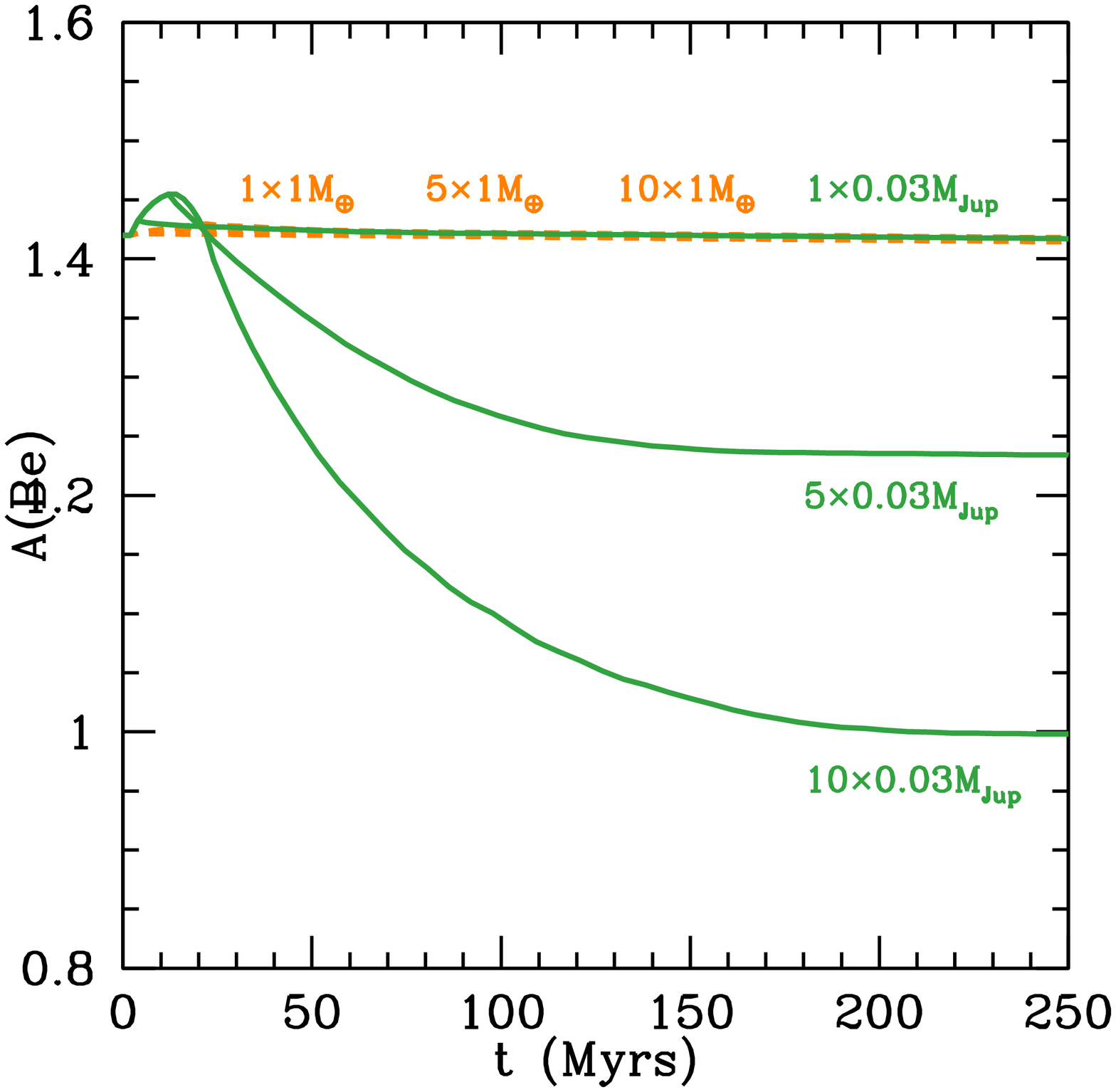}
\caption{Variations of Li and Be surface abundances along the accretion/mixing period for several 1.10M$_{\odot}$ models.  The initial Li and Be abundances are taken from \citet{Grevesse93} and respectively equal to 3.31 and 1.42. The curves are labelled according to the number of accretion episodes and the accretion rate undergone.}
\label{libesurf}
\end{figure*}

Figure \ref{libesurf} shows the evolution of the Li and Be surface abundances over the accretion/mixing period. The results demonstrate that the accretion of planetesimals may decrease the surface Li and Be abundances, with a depletion rate depending on the accretion rate and the number of impact events (the Li and Be depletion increasing with the accretion rate and/or the number of impacts). The Be nuclear depleted zone being deeper that the Li one, the obtained Be destructions are smaller than the Li ones. 

\subsection{Influence of the stellar mass and the initial metal content}
In this section we present models with various masses and initial metallicities.  All the presented models undergo 5 accretion episodes of 0.03M$_{\rm Jup}$ following the same scheme as previously discussed (same time intervalle between the ZAMS and the first impact and then between successive accretion episodes). As in previous sections, the thermohaline mixing is still computed using the KRT coefficient with $C_t=12$.

\begin{figure*}
\center
\includegraphics[width=0.5\textwidth]{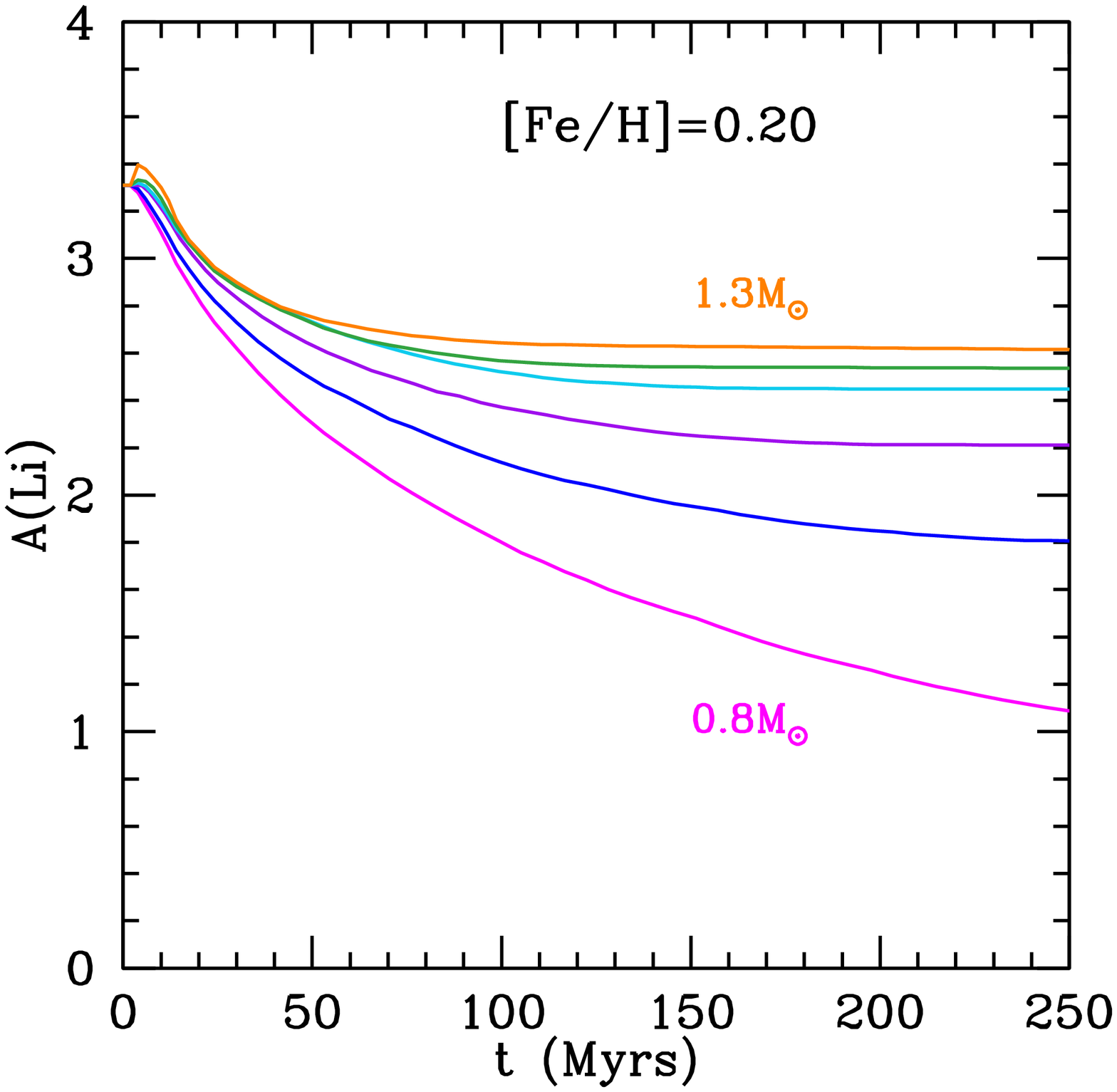}%
\includegraphics[width=0.5\textwidth]{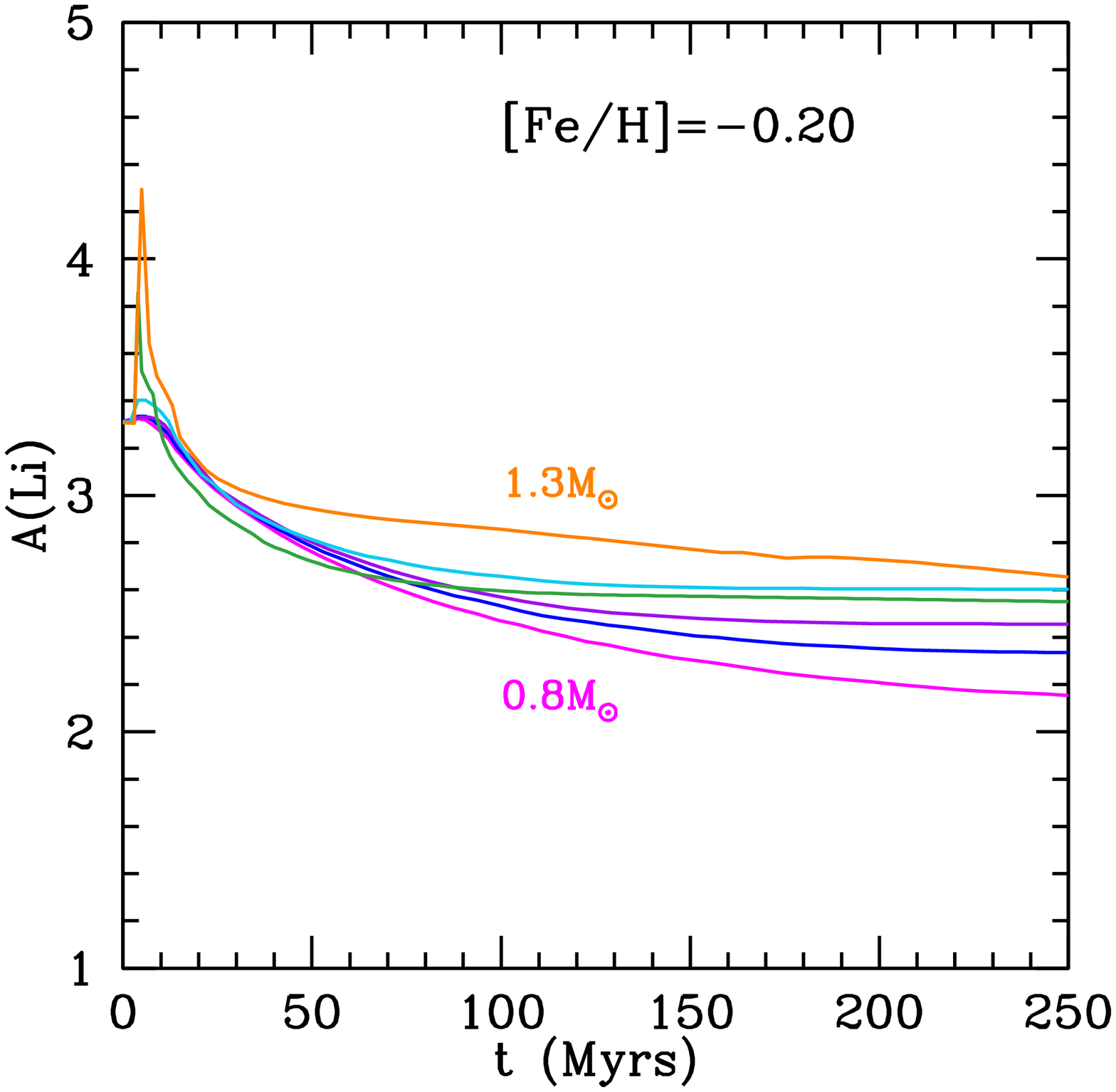}
\caption{Lithium surface abundance over the accretion/mixing period in models experiencing 5 accretion episodes of 0.03M$_{\rm Jup}$. The presented models have different masses (0.8, 0.9, 1.0, 1.1, 1.2 and 1.3M$_{\odot}$) and initial metallicities. These models have been computed with the KRT coefficient, $C_t$=12.}
\label{lisurfz}
\end{figure*}

Figure \ref{lisurfz} presents the surface Li abundance during the accretion/mixing period in two sets of models computed with different initial metal contents. For each metallicity several masses are presented. For a given metallicity, the Li-depletion induced by the combined effects of accretion and mixing decreases with increasing mass. Several competing effects lead to this result. 

The mass of the surface convective zone decreases as the stellar mass increases, as a result the dilution of a metal rich planetesimal in the external convective zone should lead to larger unstable $\mu$-gradients and then to more efficient mixing and larger Li-depletion in more massive stars. However other competing effects lead to a different result. 
The Li amount brought through the ingestion of a metal-rich planetesimal increases the Li surface abundance of the target star. This increase is more significant in more massive stars (due to their narrow convective zone). As a result more massive stars enter the mixing period with higher Li surface abundance. Moreover as the mass increases the top of the Li-depleted region moves away from the bottom of the convective zone. Larger mixed regions are then needed to connect the surface and the Li-nuclear burning zone. Both these effects (the accretion-induced Li increase and the growing distance between the convective zone and the Li burning region) make the Li depletion smaller in more massive stars. 

\begin{figure*}
\center
\includegraphics[width=0.5\textwidth]{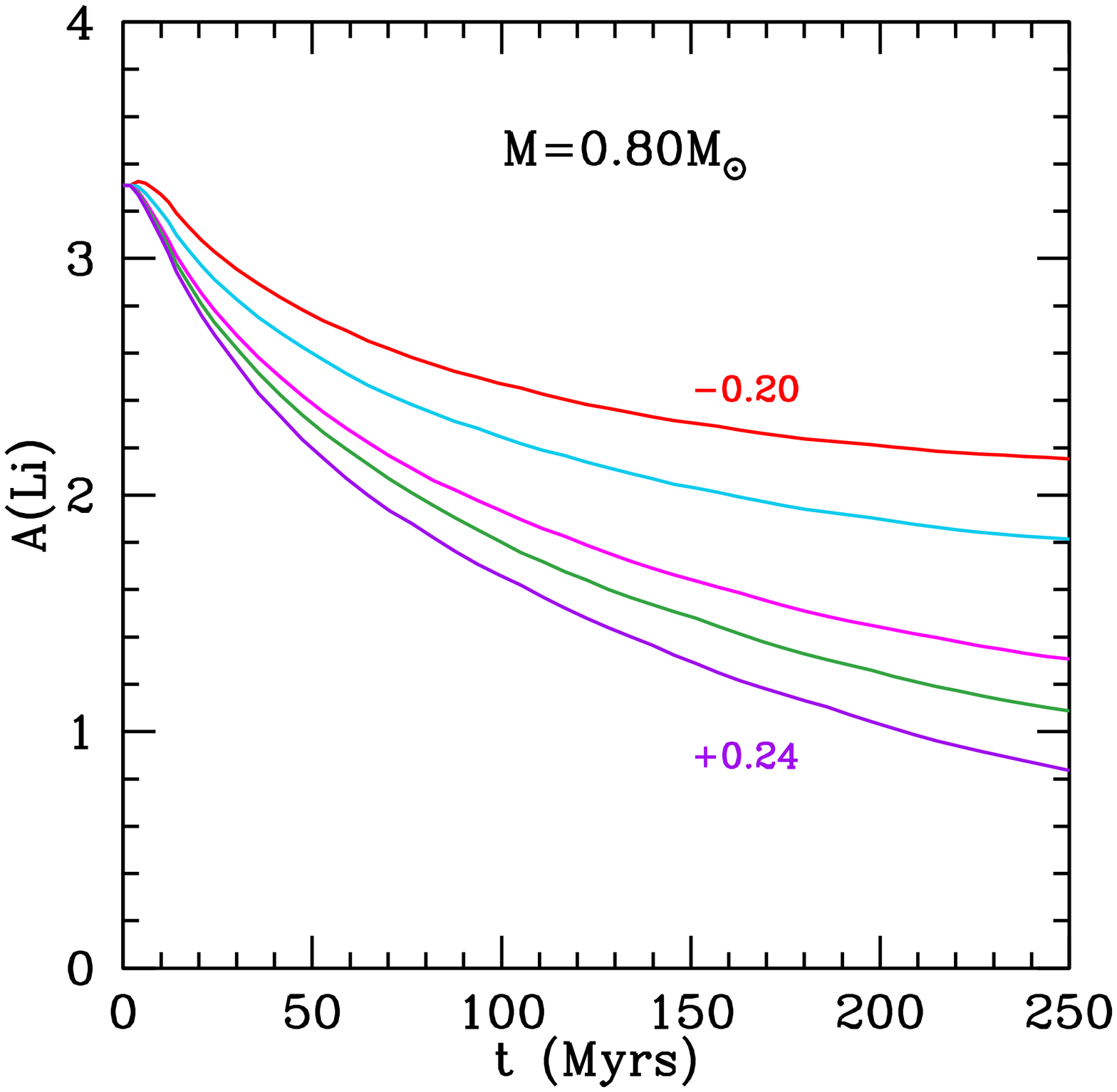}%
\includegraphics[width=0.5\textwidth]{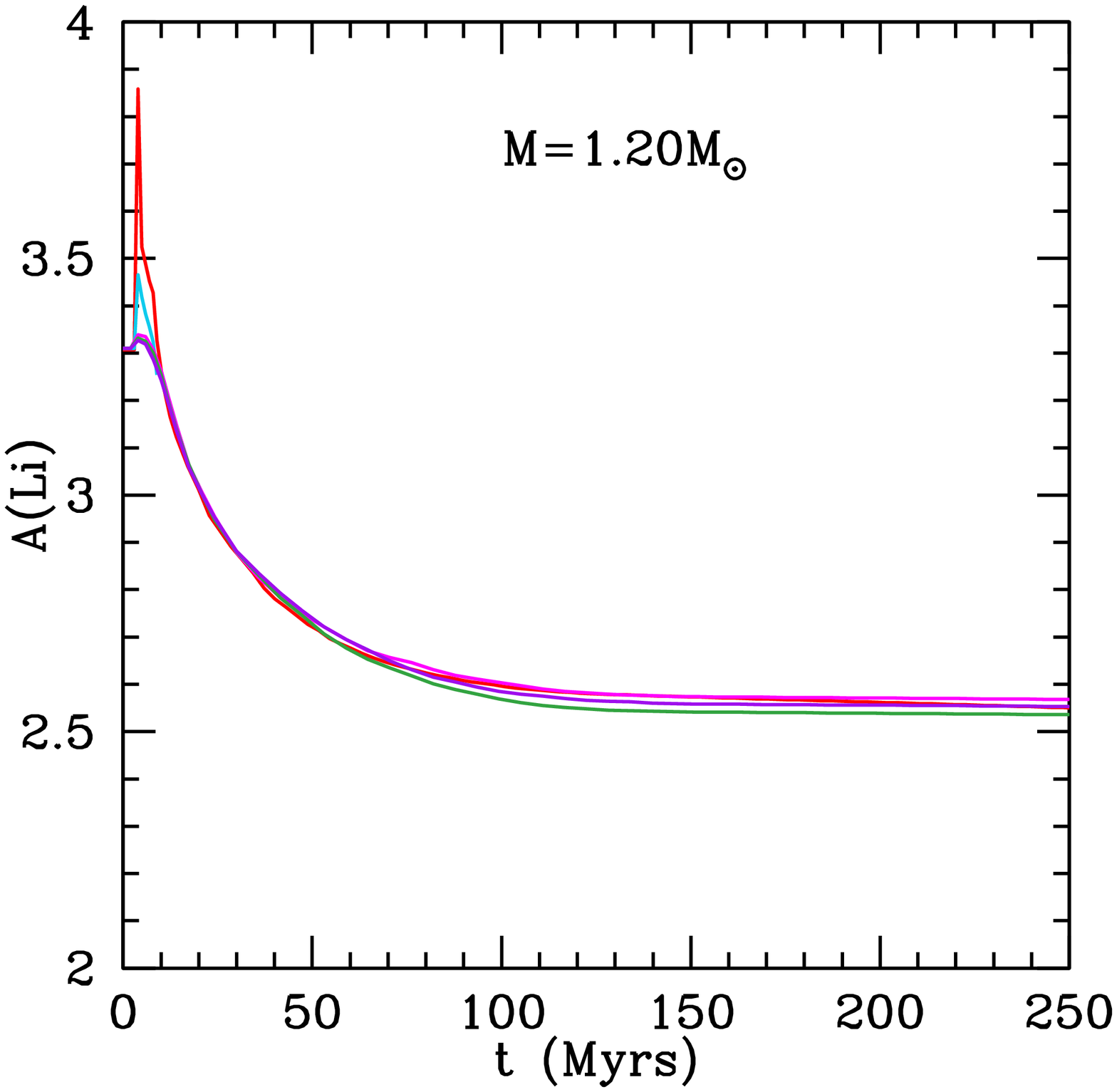}
\caption{Lithium surface abundances over the accretion/mixing period in models experiencing 5 accretion episodes of 0.03M$_{\rm Jup}$. The presented models have different initial metallicities ([Fe/H]=-0.20, 0.00, 0.16, 0.20, 0.24) and different masses (KRT coefficient, $C_t$=12). For the 1.20M$_{\odot}$ models, the five curves merge.}
\label{mz1}
\end{figure*}

\begin{figure}
\center
\includegraphics[width=0.03\textwidth,bb=4 50 50 550]{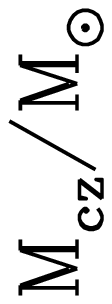}%
\includegraphics[width=0.47\textwidth]{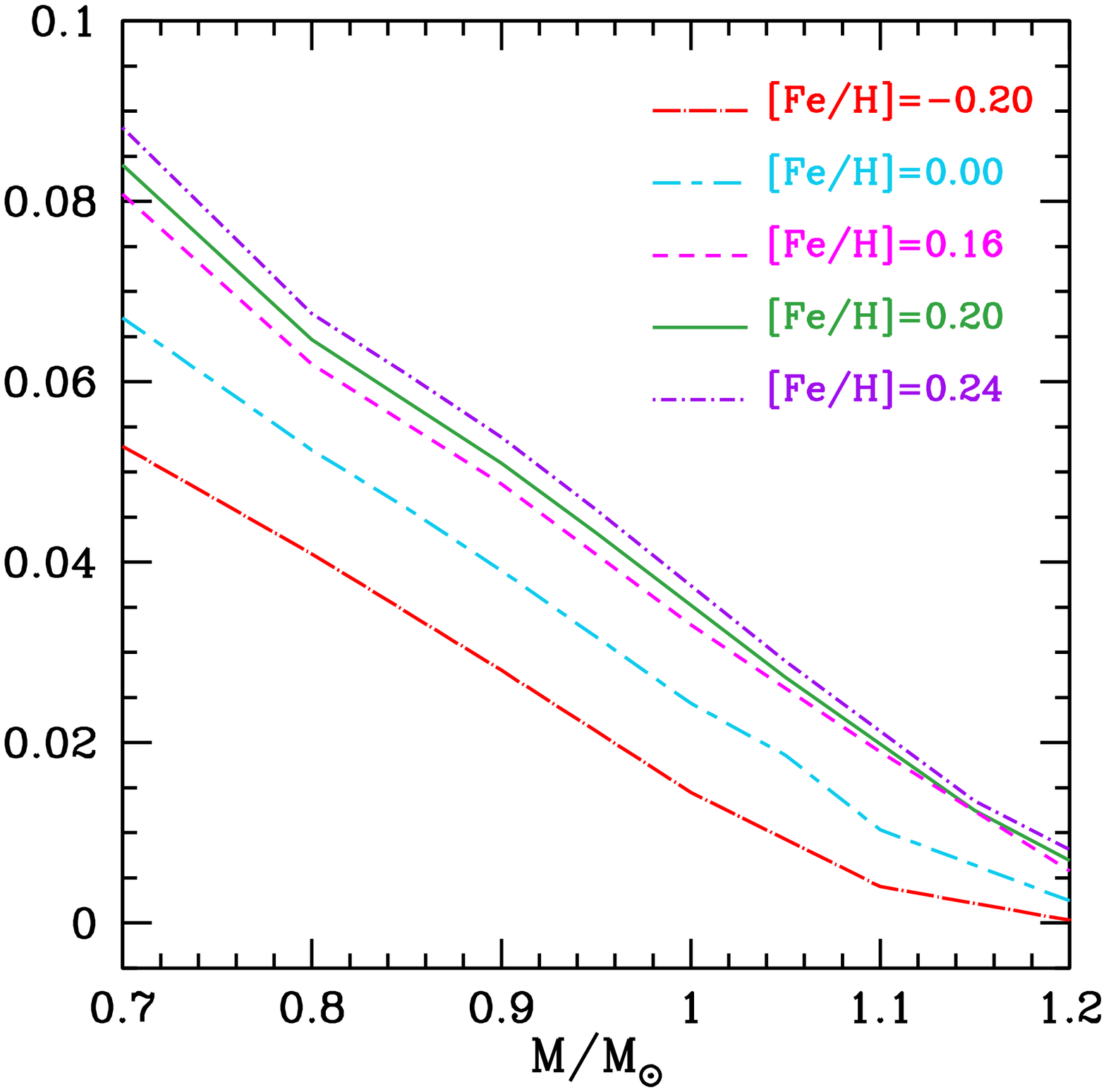}
\caption{Surface convective mass as a function of the stellar mass in ZAMS models. }
\label{mmec}
\end{figure}

Figures \ref{mz1}a and b compare the Li depletion in models with the same mass but with different metallicities. It shows that the accretion/mixing induced Li depletion increases with the initial metal content. For a better understanding of this result we present on Figure \ref{mmec} the surface convective mass as a function of the stellar mass for ZAMS models with various metallicities. It shows that the mass of the surface convective zone increases with metallicity. This feature confirms that several competing effects leads to larger depletion in stars with larger surface convective zones.

\subsection{Results obtained with the KRT coefficient and $C_t=1000$}
Up to now, we presented results obtained using the prescription proposed by KRT with the adjustment $C_t=12$. As already discussed, this coefficient is in much better agreement with the results of recent numerical computations \citep{Traxler11} than the one proposed by \citet{Charbonnel07}, using $C_t=1000$. We nevertheless computed models using this large value of $C_t$ and give the results below for comparison. 

\begin{figure*}
  \centering
  \begin{minipage}[c]{0.5\textwidth}
    \centering
    \includegraphics[width=\textwidth,bb=240 162 565 690]{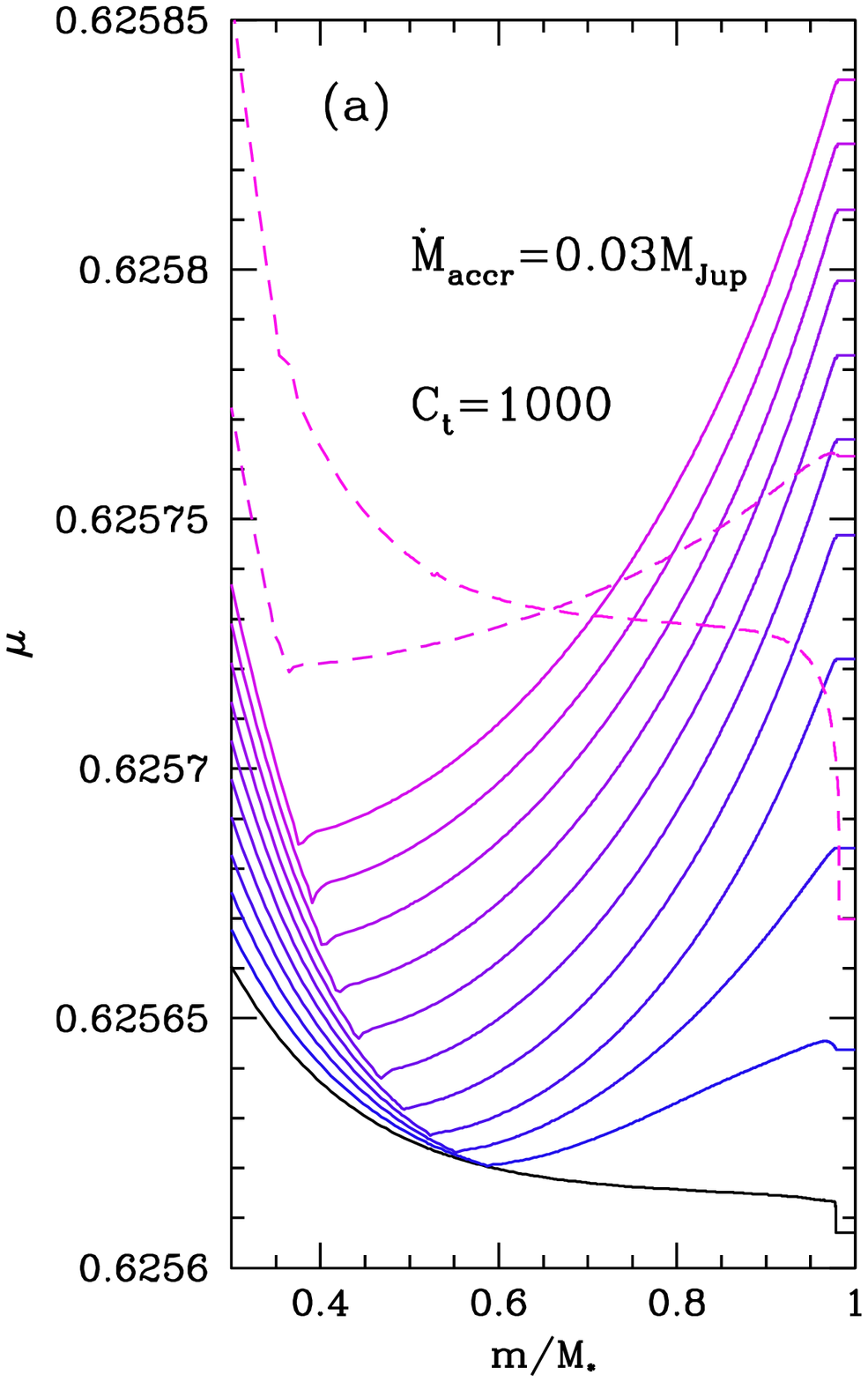} %
  \end{minipage}%
  \begin{minipage}[c]{0.5\textwidth}
    \centering
    \includegraphics[width=0.81\textwidth,bb=315 430 565 690]{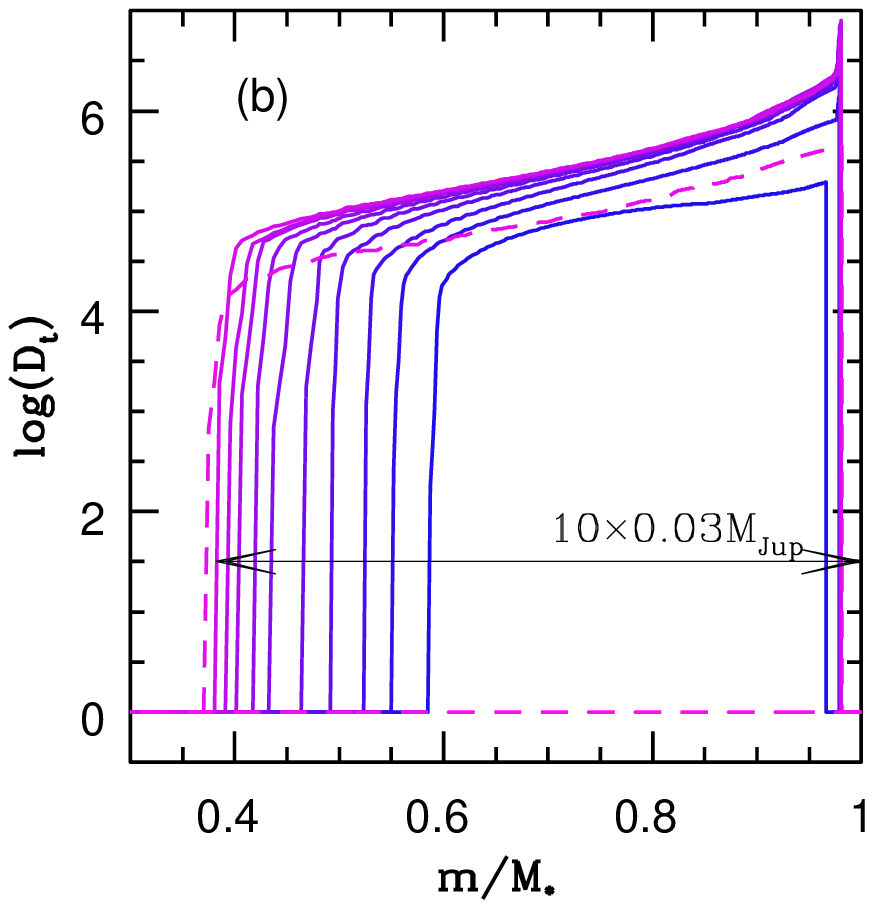} %
    \includegraphics[width=0.81\textwidth,bb=315 430 565 690]{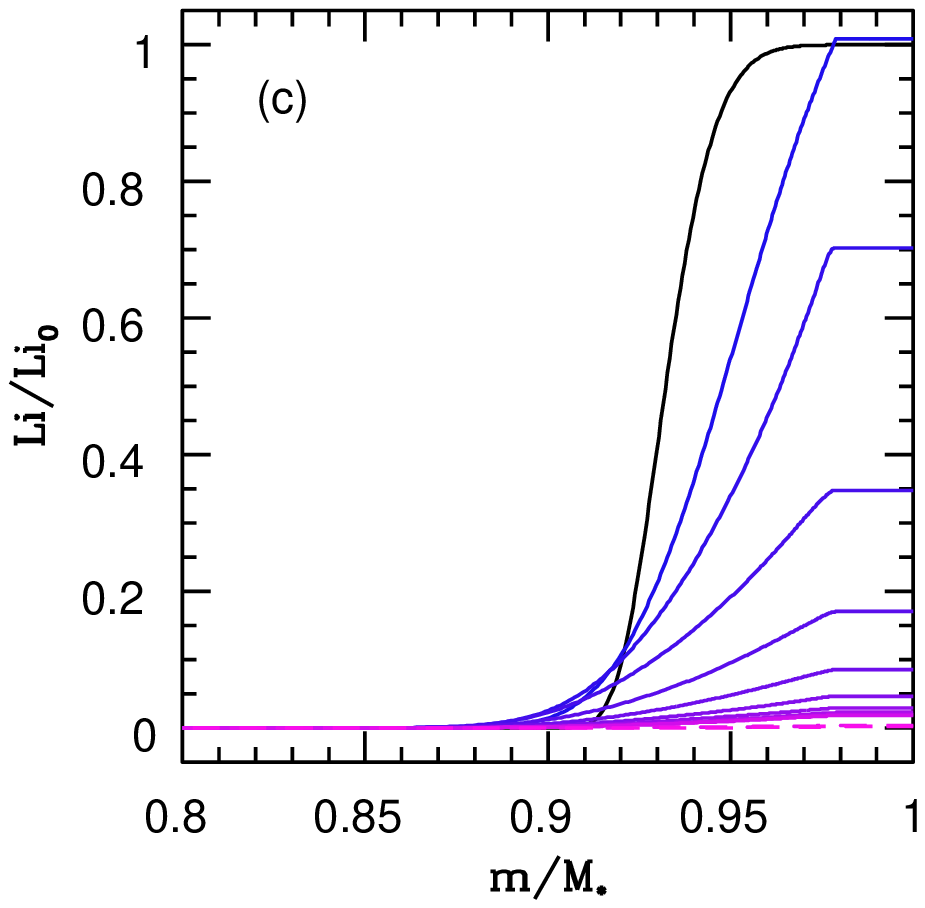} %
  \end{minipage}%
\caption{Same figure as Figure \ref{dkip100b} for a  1.10M$_{\odot}$ models experiencing 10 accretion events of 0.03M$_{\rm Jup}$, computed with $C_t$=1000. The line convention is the same as on Figure \ref{dkip100b} except for the light magenta dashed lines which represent the profiles at 30 and 51 Myr on the ZAMS (from top to bottom). The black arrow on the upper right panel shows the extension of the mixed zone after 10 accretion episodes. A small stable helium zone develops after one accretion event and never come back afterwards.}
\label{dkip1a}
\end{figure*}

Figure \ref{dkip1a} present the results obtained with $C_t=1000$ for a 1.10M$_{\odot}$ model (with Y$_0$=0.29 and [Fe/H]=0.20 as previously presented models), undergoing ten accretion episodes of 0.03M$_{\rm Jup}$. The accretion sequence is the same as previously : it starts 2 Myr after the ZAMS beginning and the successive accretion events are separated by 2 Myr.

The models presented on Figures \ref{dkip100b} and \ref{dkip1a} differ only through the value used for the $ C_t$ factor. A larger $ C_t$ value increases the diffusion coefficient and consequently produces a larger mixing. The 2 Myr-mixing occurring in-between two successive accretion events is more efficient in smoothing the unstable stratification. Consequently the succession of several accretion episodes produces smaller unstable gradients in models with larger $ C_t$ values. This feature is indeed observed (cf. Figures \ref{dkip100b}a and \ref{dkip1a}a). A retroactive effect of smaller $\mu$-gradients is to decrease the diffusion coefficient (cf. Eq. \ref{dth}). The mixing efficiency then results from the competition between two opposite effects. The larger $C_t$ tends to increase the diffusion coefficient and the mixing efficientcy, thereby leading to smaller $\mu$-gradient, but on the other hand these smaller unstable gradients tend to decrease it. 

\begin{figure}
\center
\includegraphics[width=0.02\textwidth,bb=17 50 50 550]{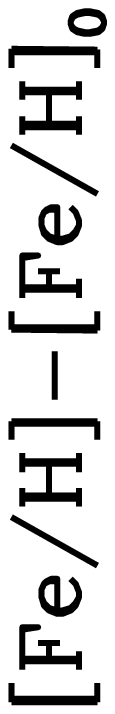}%
\includegraphics[width=0.48\textwidth]{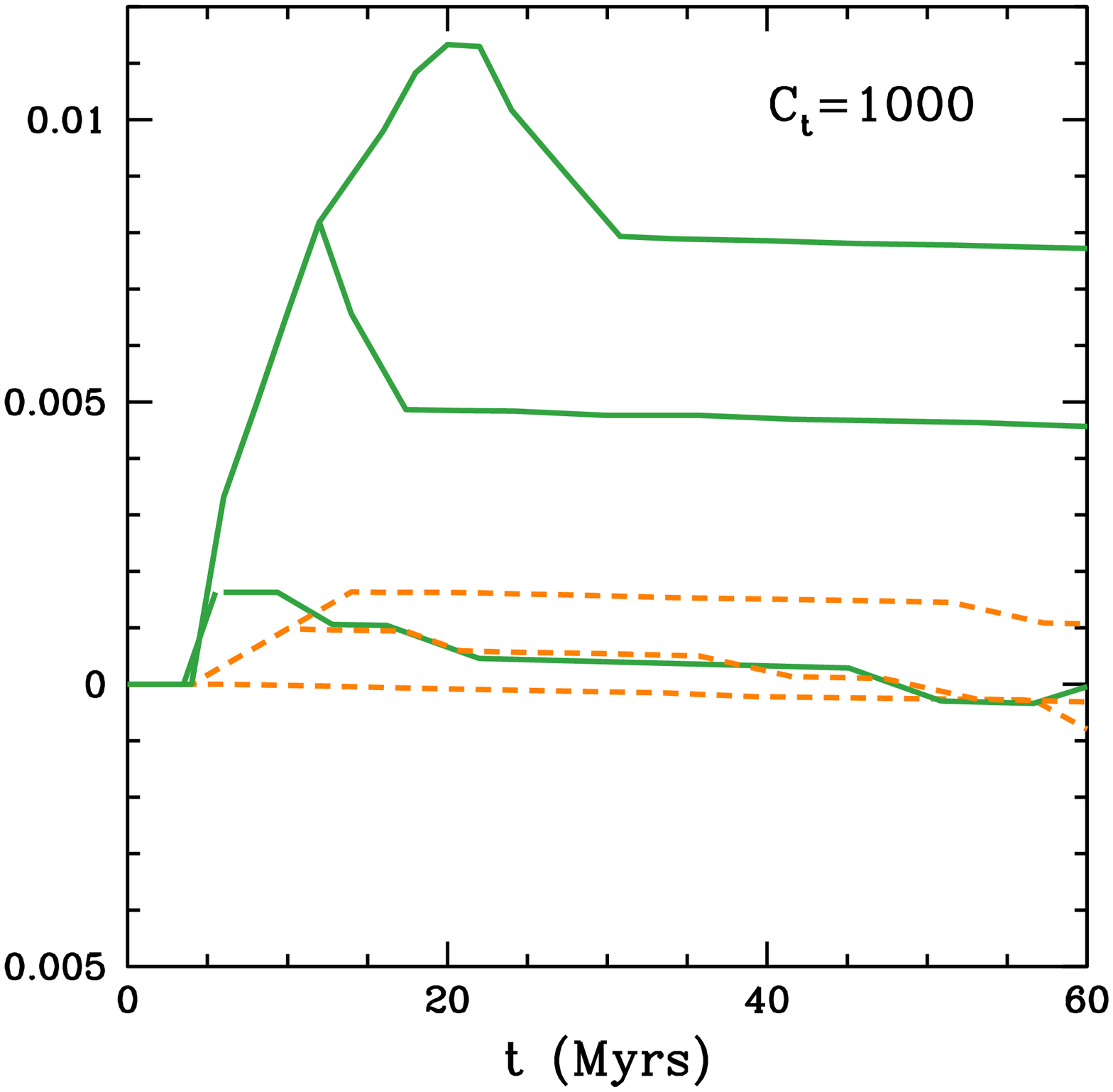} %
\caption{Same figure as Figure \ref{fesurh} for models computed with $C_t=1000$. Dashed lines are models experiencing (from bottom to top) : 1, 5 or 10 accretion events of 1M$_{\oplus}$. Solid lines are models experiencing (from bottom to top) : 1, 5 or 10 accretion events of 0.03M$_{\rm Jup}$.}
\label{dkip1fesurh}
\end{figure}
\begin{figure*}
\center
\includegraphics[width=0.5\textwidth]{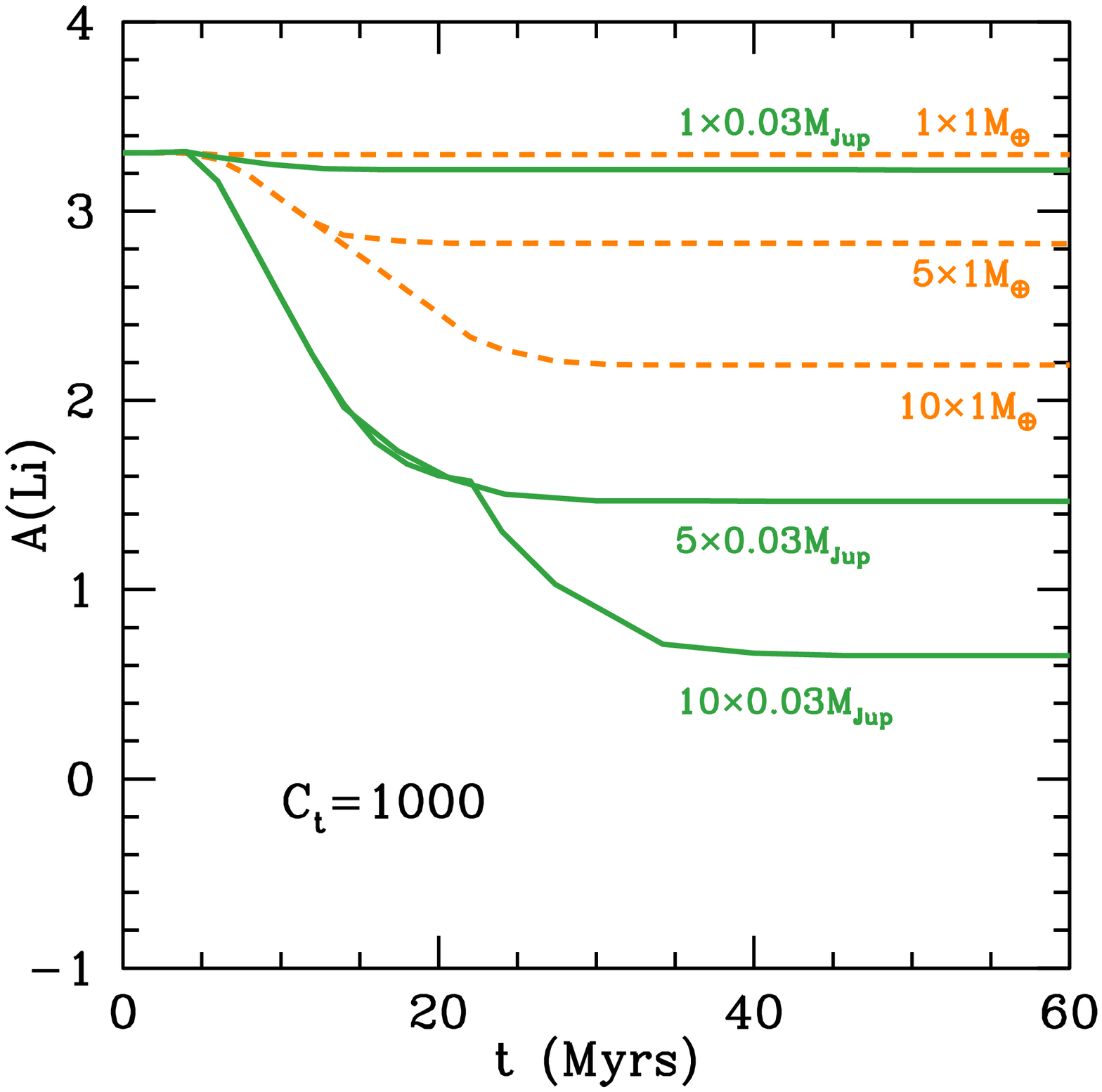}%
\includegraphics[width=0.5\textwidth]{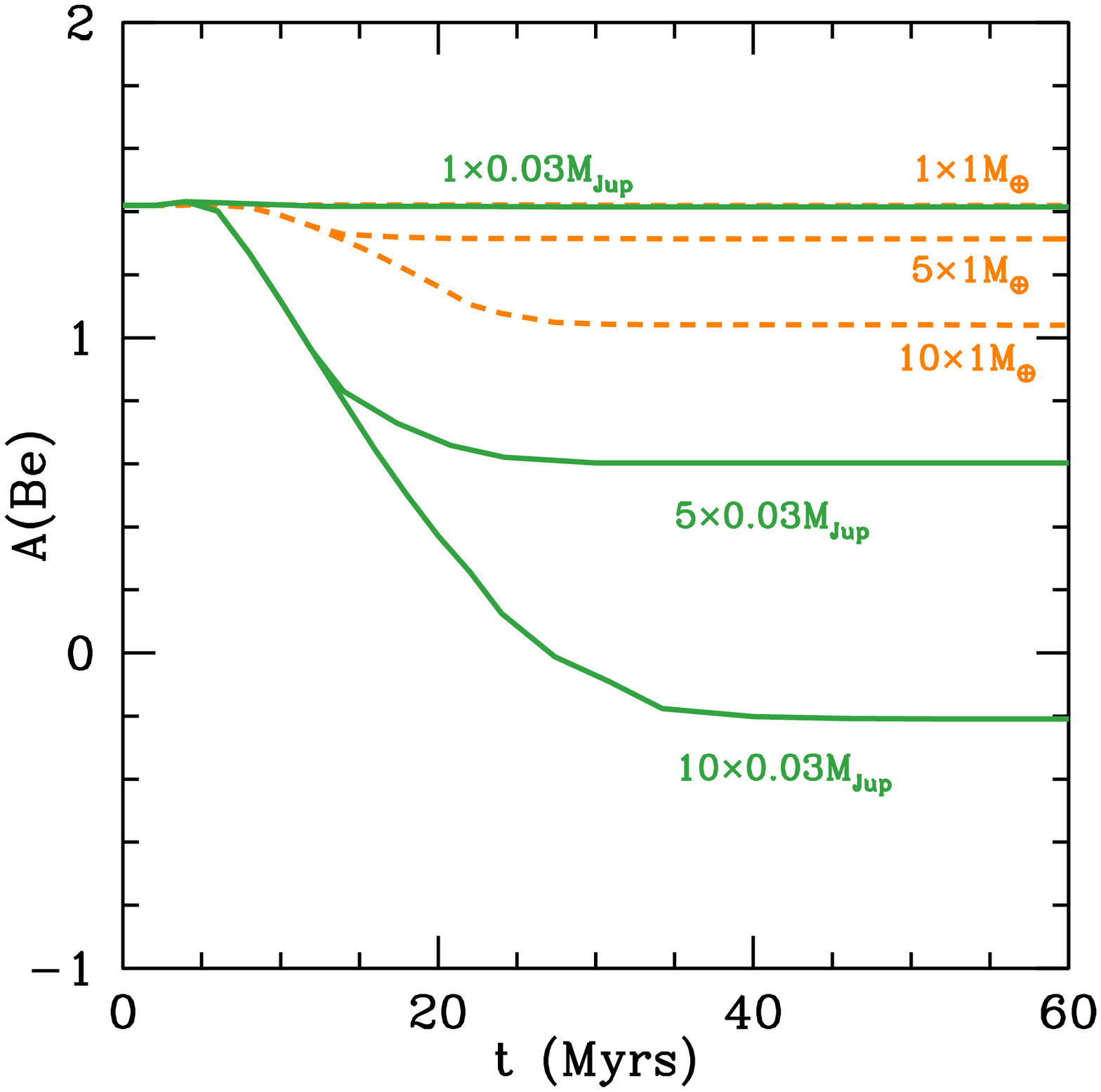}%
\caption{Same figure as Figure \ref{libesurf} for models computed with $C_t=1000$. The line convention is the same as in Figure \ref{libesurf}.}
\label{dkip1blibesurf}
\end{figure*}

Figure \ref{dkip1fesurh} illustrates the variations of the accretion-induced overmetallicity during the accretion/mixing perio, using the large diffusion coefficient (for comparison with Figure 4). Figure \ref{dkip1blibesurf} shows the lithium and beryllium depletion due to accretion and mixing. As previously discussed they depend on the accretion rate and the number of accretion events but for a given accretion scenario, the obtained destructions are larger than those obtained with a smaller diffusion coefficient (cf. Figure \ref{libesurf}).

\subsection{Influence of the accretion events frequency
}
\label{dtaccr}
In previous sections, we discussed either the cases of single accretion events, or the cases of several accretion events occuring every 2 Myr. When a new event occurs, the element abundances are first modified inside the convective zone due to the chemical composition of the accreted matter, and then they go on varying due to the thermohaline mixing. In the case of lithium, each new event leads to a rapid lithium increase followed by a slower decrease induced by thermohaline convection and nuclear burning. Obviously the final results depend on the amount of accreted matter but also on the time scale between two events. If another accretion occurs before the complete settling of the previous one, the effects overlap.

\begin{figure*}
\center
\includegraphics[width=0.4\textwidth,bb=230 150 575 700]{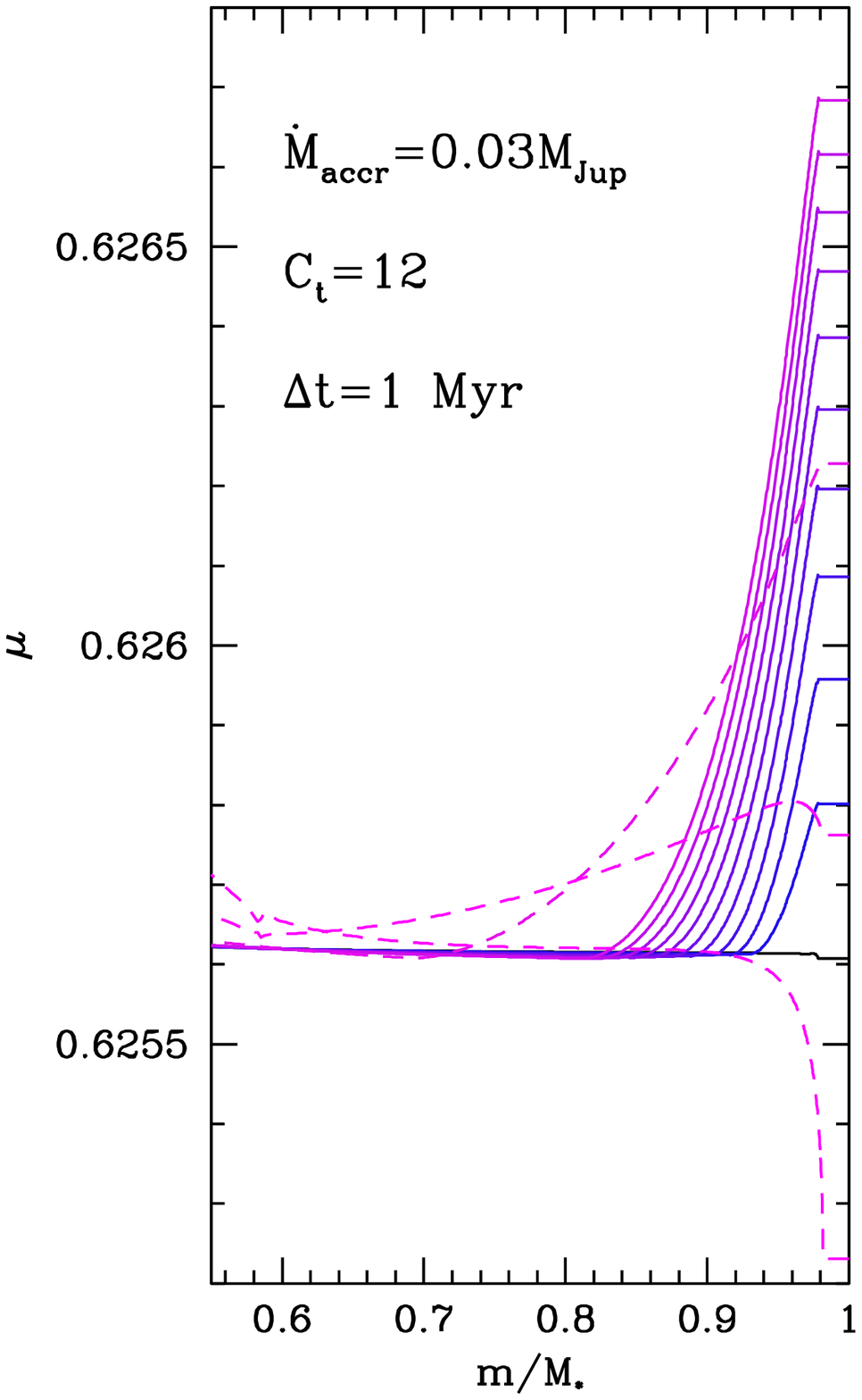}%
\includegraphics[width=0.4\textwidth,bb=230 150 575 700]{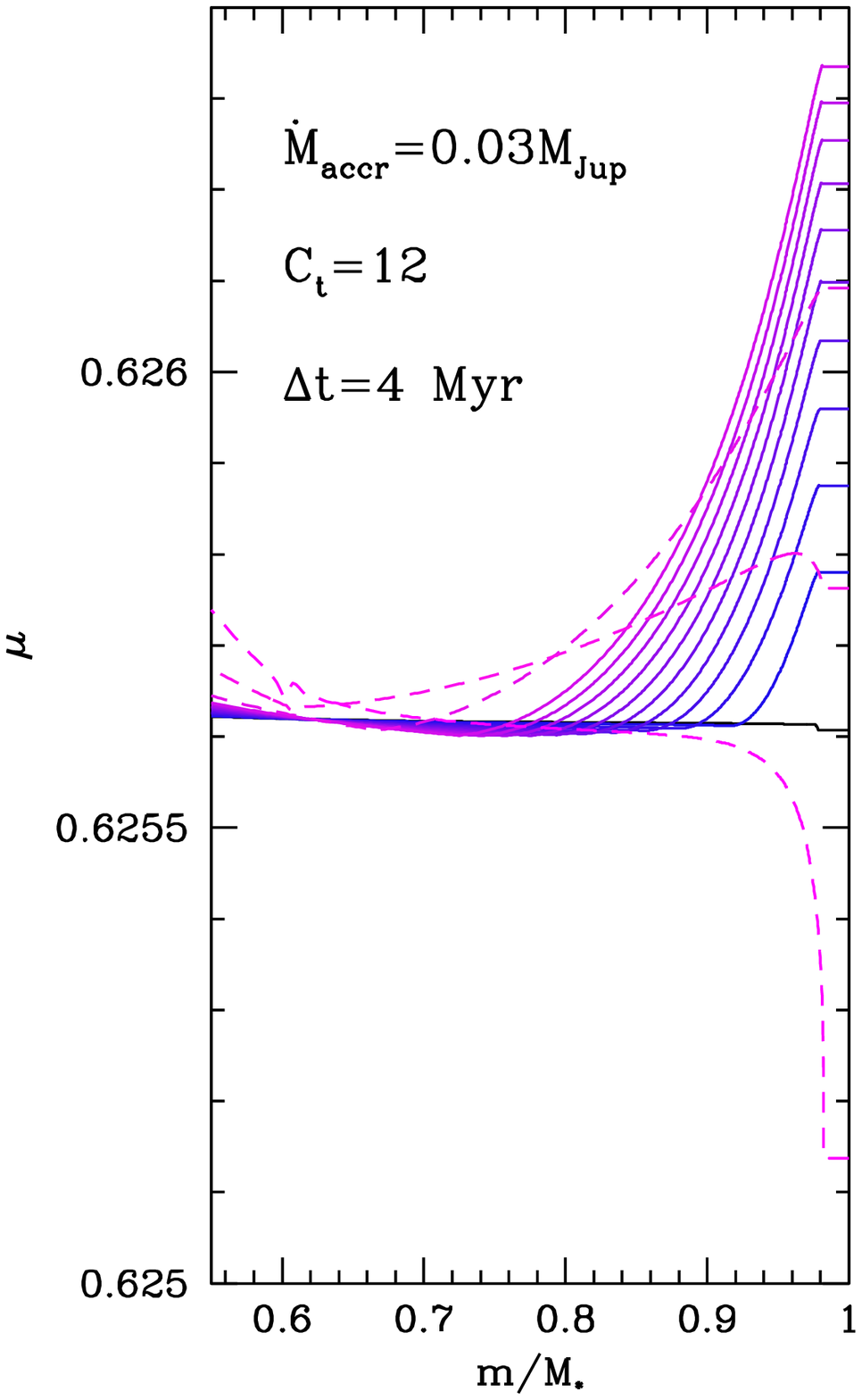} %
\includegraphics[width=0.4\textwidth,bb=310 430 575 700]{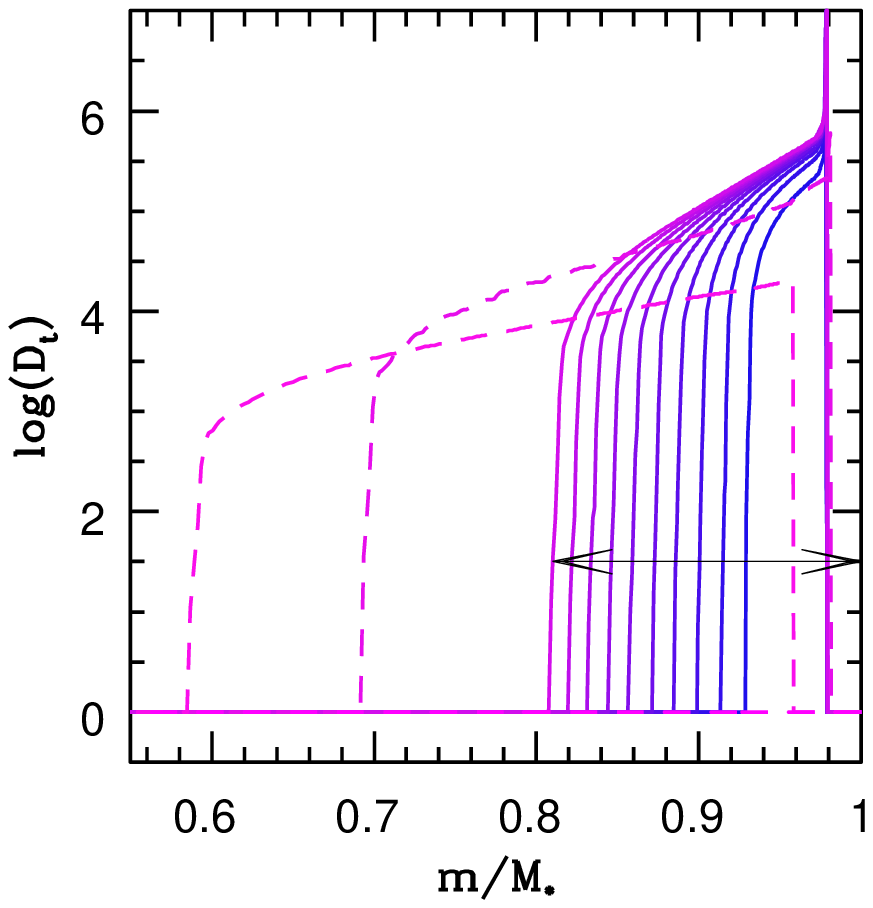}%
\includegraphics[width=0.4\textwidth,bb=310 430 575 700]{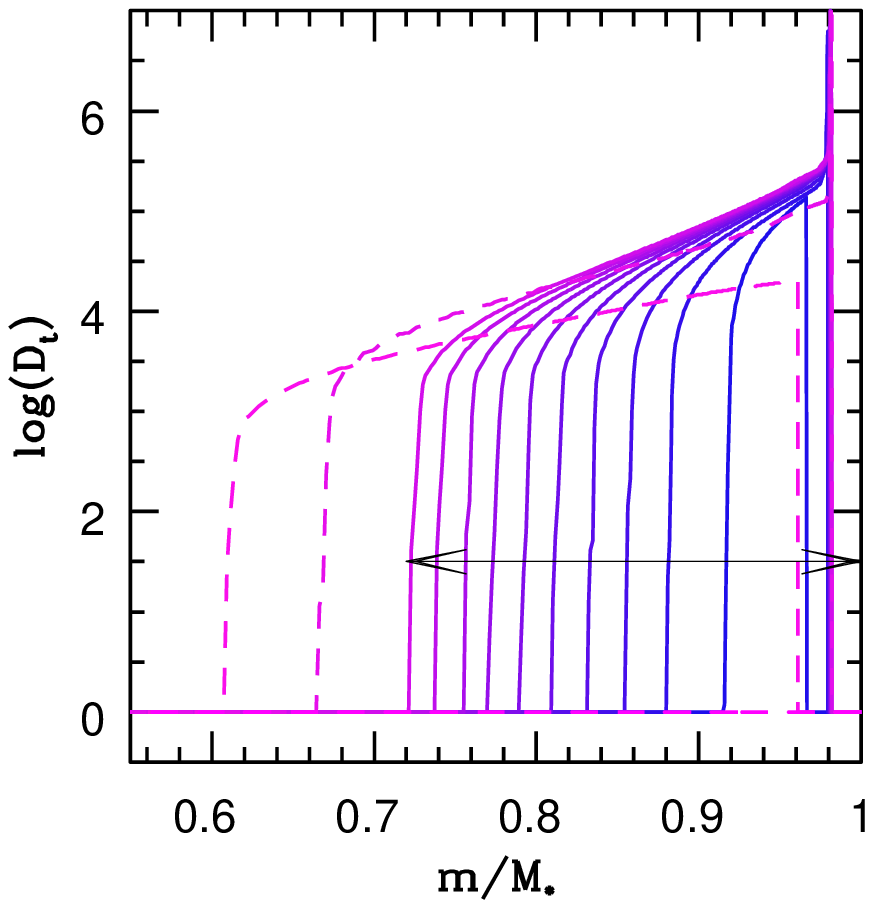}%
\caption{Molecular weight and diffusion coefficients during the accretion/mixing process in two 1.10M$_{\odot}$ models experiencing 10 accretion episodes of 0.03M$_{\rm Jup}$ with timestep of 1 or 4 Myr. The black lines represent the $\mu$-profiles at 2Myr on the ZAMS, just before the first accretion event. The dark blue to magenta solid lines present the profiles after each new accretion event. The light magenta dashed lines show profiles when the accretion has stopped but when the thermohaline mixing still proceeds : at 26, 107 and 229 Myr for the left panel, at 57, 122 and 260 Myr for the right panel. The horizontal arrows in the lower panels locate the mixed region after 10 accretion events.}
\label{mumdkipdt}
\end{figure*}

\begin{figure*}
\center
\includegraphics[width=0.5\textwidth]{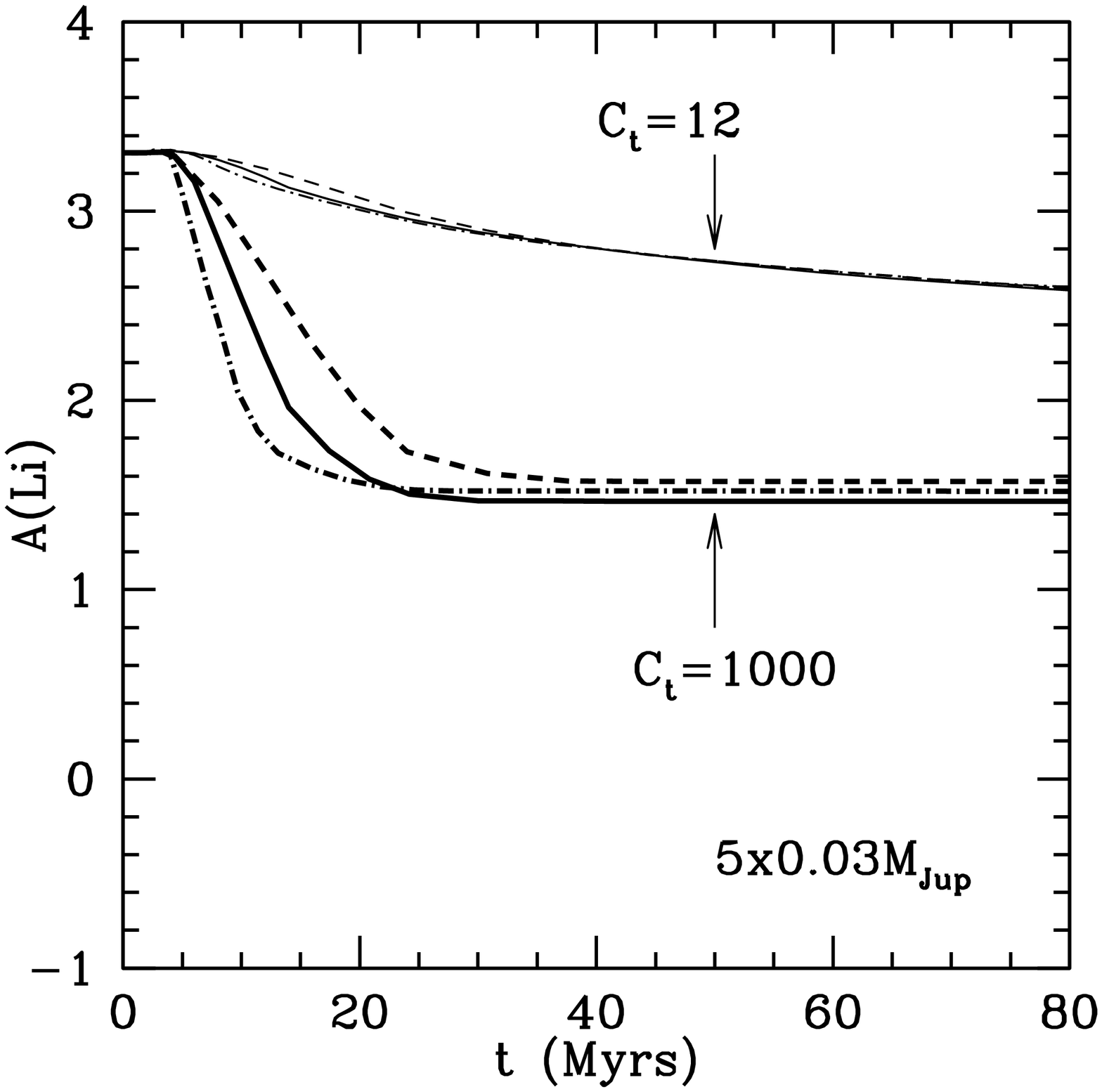}%
\includegraphics[width=0.5\textwidth]{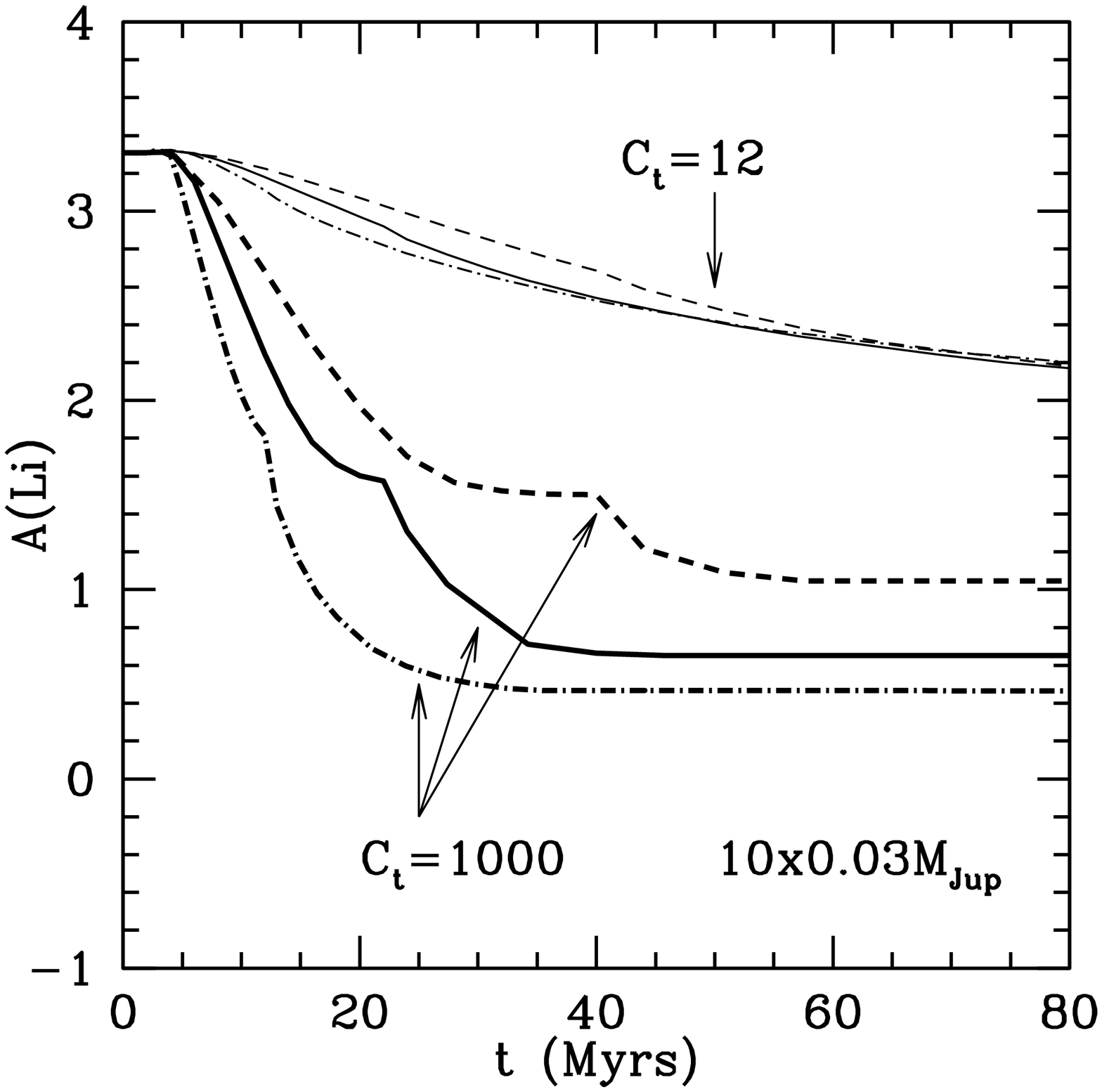}%
\caption{Lithium depletion in models computed with different time intervals ($\Delta t$) between successive accretion events. Left figure : models experiencing 5 accretion events of 0.03M$_{\rm Jup}$. Right figure : models experiencing 10 accretion events of 0.03M$_{\rm Jup}$. For each accretion scenario, two series of models are computed : one with C$_t$=12, one with C$_t$=1000. For each C$_t$ value three models are presented assuming different $\Delta t$ values :  1 Myr (dashed-dotted lines), 2 Myr (solid lines), 4 Myr (dashed lines).}
\label{lisurfdt}
\end{figure*}

In this section, we discuss the effects on lithium of changing the time interval between successive accretion events. We have tested, for accretion amounts of 0.03M$_{\rm Jup}$, the cases of events separated by 1 Myr and 4 Myr. The results are presented in Figure \ref{mumdkipdt} for comparison with Figure \ref{dkip100b}. Figure \ref{lisurfdt} shows the surface lithium abundance variation with time, together with the previously studied case of 2 Myr. The left graph displays the results for five events and the right graph for ten events. In each case, two series of models are computed : one with  $C_t$=12 and the other one with C$_t$=1000. 

The final results are not trivial. For larger time intervals, the $\mu$-gradients before every new accretion event are smaller, so that the diffusion coefficients and thus the mixing efficiencies are also smaller, leading to less lithium destruction. On the other hand, the total mixing episodes last longer, in favor of more lithium destruction.  Figure \ref{lisurfdt} shows that these two combined effects can lead to opposite results according to the circumstances. For $C_t$=12, the results obtained with the three different time intervals are very similar, although lithium is slightly less depleted for increasing intervals. For $C_t$=1000, the situation is different if accretion stops after five or ten events. For five events, the results are also very similar but this time the case of events separated by 1 Myr lead to slightly less depletion than the other cases. On the other hand, for ten events, the situation is similar to the case described for $C_t$=12, except that the difference is more important. An interesting point in this case is that the end of the accretion period is clearly visible in the graph. It corresponds to the time when no fresh lithium is brought anymore, so that the lithium abundance variation is only due to nuclear destruction.

\subsection{Computations using recent diffusion coefficients deduced from numerical simulations
}

As already mentionned, the expressions for the diffusion coefficients deduced from recent numerical simulations lead to results close to the KRT ones with $C_t = 12$, except for the boundaries of the thermohaline regime. Here we are basically interested in the $R_0 = 1$ limit, which corresponds to the boundary between the dynamical convection and thermohaline mixing. In the KRT prescription, for which $D_{th}$ becomes very large at this limit, there is a simple continuity between the two regimes. After accretion, if $R_0 < 1$, the region below the standard convective zone becomes dynamically unstable, which leads to complete mixing and may be modelled by a very large diffusion coefficient. As the $\mu$-gradient decreases, $R_0 $ increases. When it reaches unity, thermohaline convection begins. One thus expects a continuity between the diffusion coefficients, which means that when $R_0 $ becomes slightly larger than one, $D_{th}$ should rapidly decrease from a very large value to the value obtained in the intermediate regime. This description is well fitted by the KRT coefficient, but not by the TGS one where $D_{th}$ remains finite as it reaches unity. However, a close look at their Figure 2, which presents the turbulent heat and compositional fluxes as a function of their rescaled density ratio $r$, reveals a clear tendency for the numerical simulation points to go up to large values for $r = 0$, corresponding to $R_0 = 1$. We thus infer that their should be a small transition region between the TGS value and the very large one needed for $R_0 < 1$, consistently with the simulations. On the other hand, the value of the mixing coefficient as given by TGS drops to zero at the other limit, for $r = 1$, corresponding to $R_0 = 1/\tau$. This feature, which is not present in the KRT coefficient, is consistent with the fact that  $D_{th}$ must become zero for $R_0 > 1/\tau$.

\begin{figure}
\center
\includegraphics[width=0.5\textwidth]{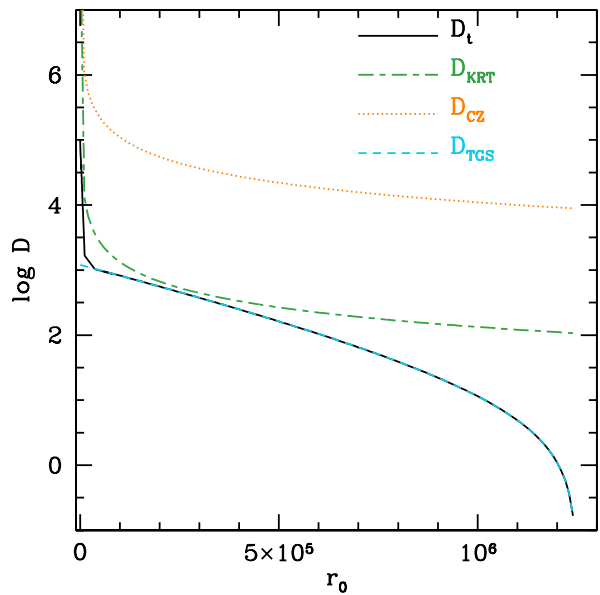}%
\caption{Thermohaline diffusion coefficients (cm$^2$.s$^{-1}$) computed in a 1.10M$_{\odot}$ model with [Fe/H]=0.20 at 4 Myr. Long dashed-small dashed (green) line: KRT coefficient; dashed (blue) line: TGS coefficient; solid (black) line: TGS coefficient including a transition region for $R_0$ close to one; dotted line: coefficient used by \cite{Charbonnel07}.}
\label{coefficients}
\end{figure}

Figure \ref{coefficients} displays the KRT and the TGS coefficients precisely computed as a function of $R_0$ in one of our models (M=1.10M$_{\odot}$, [Fe/H]=0.20 at 4 Myr). This graph is similar to Figure 3 of \citet{Traxler11}. The difference comes from the fact that TGS used constant $\kappa_{\mu}$, $\kappa_T$ and $\nu$ in their computations, whereas we use the exact values computed in our model. These values vary with depth and with age, as well as the $\mu$-gradient.

The two coefficients are very similar in the region between $R_0 = 5 \times 10^4$ and $R_0 = 5 \times 10^5$, i.e. between $1/50\tau$ and $1/5\tau$, with $\tau = 2.5 \times 10^6$. For $R_0$ close to one, $D_{TGS}$ has a finite value which would produce a discontinuity with the case $R_0 < 1$. For this reason, we have added a transition region in our computations (solid black line, label $D_t$). For the limit where $R_0$ becomes close to $1/\tau$, the TGS coefficient decreases rapidly, contrary to the KRT one. Also shown on this graph is the coefficient used by \citet{Charbonnel07}, labelled $D_{CZ}$.

\begin{figure*}
\center
\includegraphics[width=0.5\textwidth]{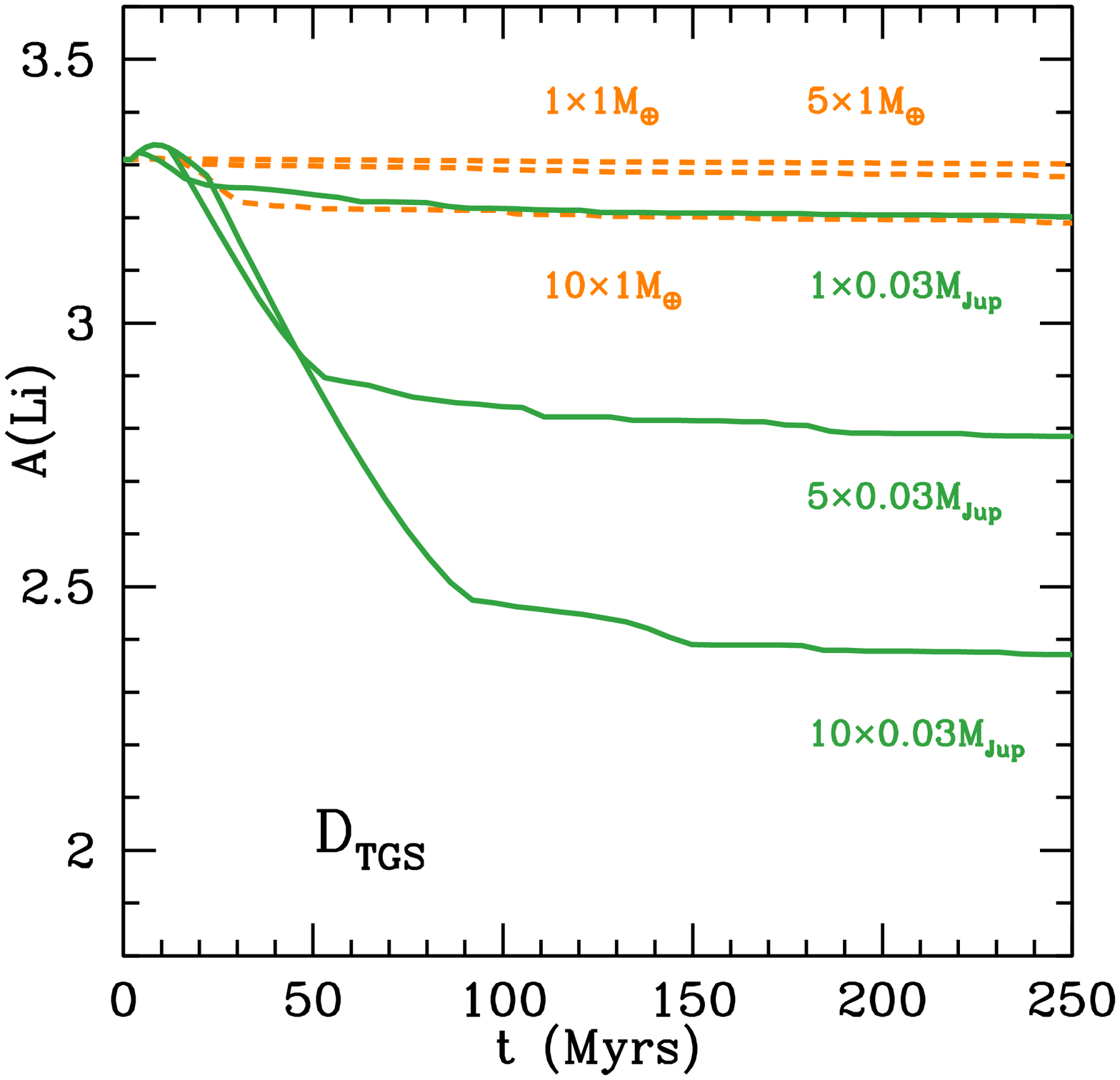}%
\includegraphics[width=0.5\textwidth]{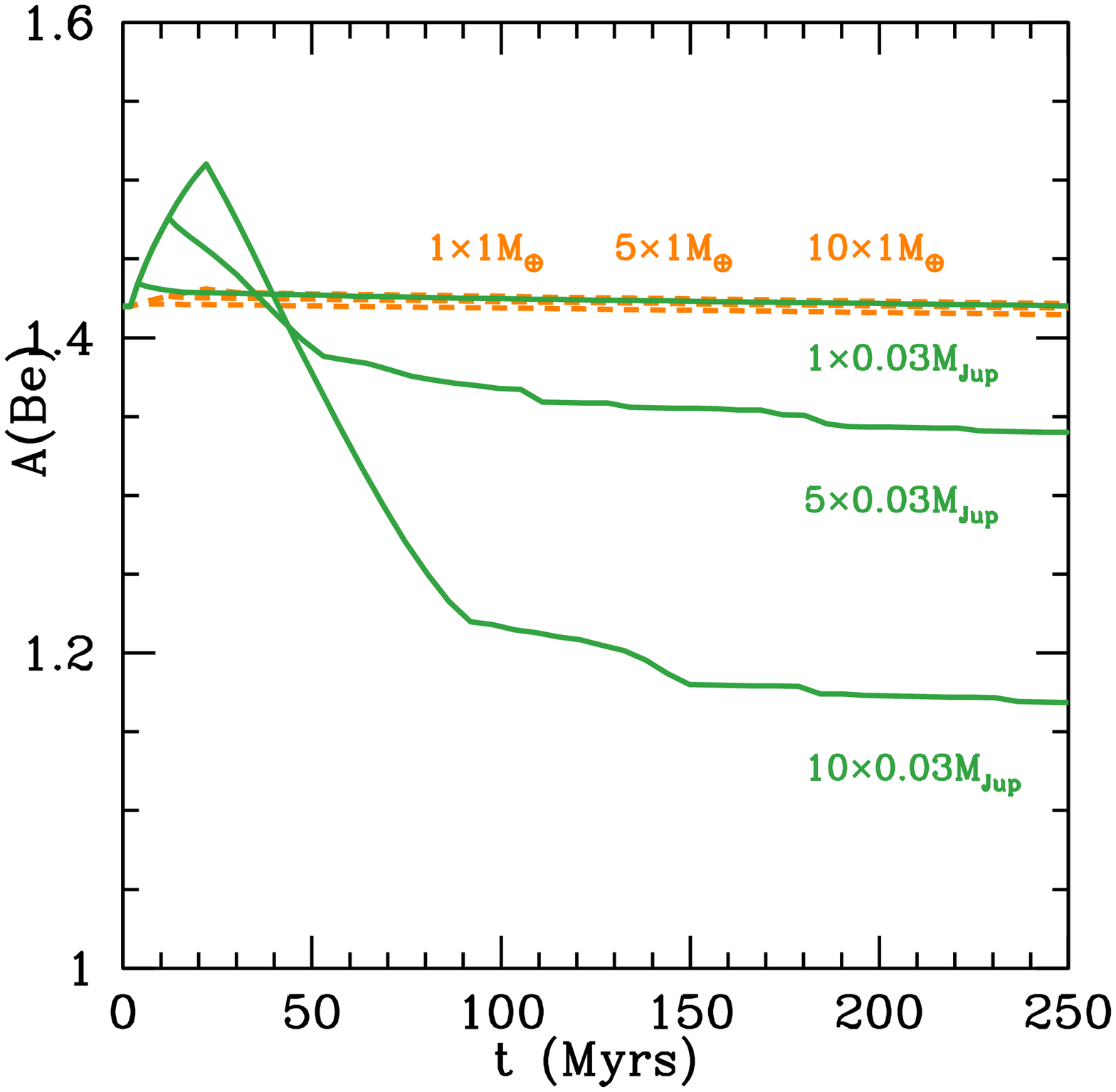}%
\caption{Same as Figures 5 with the  \citet{Traxler11} prescription for the mixing coefficient.}
\label{garaud}
\end{figure*}

Here we present results obtained with the TGS diffusion coefficient, with a very small transition region below $r = 0.02$, similar to the KRT one, to preserve the continuity with the dynamical regime. Figure \ref{garaud} displays the lithium and beryllium abundance variations with time for various accretion scenarios, which must be compared with Figure 5. We can see that the obtained lithium and beryllium depletions are very similar, although slightly smaller, that those obtained with the KRT prescription with $ C_t = 12$. In the worst case, the relative difference is of order $25\%$. On the contrary, the depletions obtained with $ C_t = 1000$ are much too large.

\section{Summary and discussion}

We have investigated the effects of planetesimal accretion onto solar-type stars. We have shown that the dilution of metal-rich material in the surface convective zone of young main sequence stars creates an unstable $\mu$-gradient at the transition between the convective and radiative zones. The induced double diffusive instabilities, the so-called thermohaline convection, result in a mixing which dilutes the metal-rich matter until the $\mu$-stratification becomes stable. The thermohaline convection effects combined with the stabilizing action of atomic diffusion softens the inverse $\mu$-gradient on a few millions or tens of million years, depending on the considered accretion scenario. Note that this time scale is much larger than the one suggested by \cite{Vauclair04}. The reason is that, following KRT, she used as a characteristic length the evaluated fingers dimension, whereas the real length to be used is that of the whole mixed region. At the end of the accretion/mixing period, only a very small metallicity increase may remain, much smaller than the average overmetallicity observed in exoplanet-host stars. This result is consistent with a primordial origin of this overmetallicity. On the other hand, as suggested by \cite{Garaud11}, the thermohaline mixing allows connecting the stellar surface with the Li-nuclear burning region, which may lead to rapid Li-depletion depending on the accretion rate and the number of impacts experienced by the star.

The Li destruction depends on many different parameters linked to the stellar structure and to the rate, frequency and amount of the accretion episodes. Concerning the stellar structure, the lithium depletion is obviously larger for deeper convective zones, which is related to the stellar mass and metallicity. It also depends on the age at which the accretion occurs, because the diffusion-induced $\mu$-gradients act as stabilizers for the thermohaline convection.  Concerning the accretion process itself, we have shown that different results are obtained not only for different accreted amounts, but also according to the way it is accreted : all at ones, or in successive episodes. Different stellar characteristics and various accretion scenarios produce, at the end of the accretion/mixing period, a significant dispersion in the Li abundances. 

The results also depend on the computational parameters used for the treatment of convection. In the present computations, the convective zone is computed whithin the framework of the mixing length theory, with a mixing length ratio adjusted on the solar models ($\alpha$ = 1.8). No overshooting nor extra-mixing except thermohaline convection is added. Thermohaline convection is treated using the diffusion approximation. 

We have tested several prescriptions for the diffusion coefficient. Recent 3D numerical simulations by \citet{Traxler11} allowed real improvements in the modelling. We have shown that the corresponding prescription for the diffusion coefficient leads to results very similar to the KRT one, provided that the aspect ratio to the fingers, that is the ratio of their length to their width, be taken equal to one as suggested by the simulations (see also \citet{Denissenkov10}). This corresonds to a numerical coefficient $C_t = 12$ in the KRT expression. As the diffusion coefficients are larger for larger $\mu$ gradients, which themselves decrease faster when the diffusion processes are stronger, the final lithium depletion is not simply related to the initial diffusion coefficient value. Here we have tested three cases: the KRT coefficient with $C_t = 12$, the same coefficient with $C_t = 1000$, as suggested by \cite{Charbonnel07} for the RGBs and the coefficient deduced from numerical simulations by TGS. We confirmed that the KRT, $C_t = 12$ coefficient gives results similar to the TGS one.

\citet{Garaud11} performed computations of thermohaline convection in a $1.4 M_{\odot}$ static model, using the TGS coefficient. No atomic diffusion was introduced in her computations. We can see from her results, and particularly from the values of the diffusion coefficient and the mean molecular weigth presented in her Figure 3, that our results and hers are similar, of the same order of magnitude. This convergence is quite encouraging.

The general trend of lithium depletion with age in various galactic clusters may be accounted for by rotation-induced mixing and/or internal waves in a somewhat satisfactory way (e.g.\cite{Theado03b,Talon05}). The large dispersion observed in stars of similar intrinsic parameters like mass, age, rotation velocity, chemical composition needs a more complicate process to be accounted for \citep{Sousa10,King10}. As discussed by \cite{Pinsonneault11}, the details of the rotation history of stars can help explaining these different features.  Here we suggest that the accretion/thermohaline mixing process may also be an important parameter in the description of the lithium dispersion in solar-type stars, and that it should be taken into account.

Comparing our results with the observed abundances of lithium and beryllium in solar type stars lead to several concluding point.

1) If one assumes that accretion of planetary matter onto solar type stars is a common phenomenon, thermohaline convection should often occur. On the other hand, if most of the lithium abundance behavior in solar type stars is well accounted for by rotation-induced mixing and related processes, as generally believed, it leads to constraints on the accretion efficiency. Using the realistic TGS or KRT thermohaline coefficients, we find that the stars must not accrete more than ten earth masses after their arrival on the Main Sequence, to keep their original lithium value within 0.1 dex. 

2) Stars which accrete more than ten earth masses are expected to be subject to stronger lithium depletion. As already pointed out by \citet{Theado10} and \citet{Garaud11}, this process could be an elegant explanation for the observations of stars in which strong lithium depletion is found. Note that this does not only apply to those stars for which planets have been detected, as stars may well host planets even if they have not yet been detected because of observational bias.

3) In any case, the observed overmetallicity in exoplanets-host stars is not due to accretion, and it is not possible either to explain any increase of lithium or of the $^6$Li/$^7$Li ratio by accretion, because of the thermohaline dilution.

4) The solar case deserves a more precise analysis. Whether accretion at the period of the late bombardment has participated in the presently observed lithium depletion or not depends on the amount of matter swallowed by the Sun, and also on the diffusion-induced helium gradient at the time of the event. 

5) Finally, a complete study of lithium depletion in solar-type stars needs a combined computation of this process with other depletion processes like rotation-induced mixing. Complete evolutionary sequences including accretion, thermohaline mixing and rotational mixing will be investigated and presented in a forthcoming paper.

\acknowledgments{We thank the referee for interesting and useful comments. SV acknowledges a grant from the Institut Universitaire de France.}

\end{document}